\documentstyle[12pt]{article}
\newcommand{\beq}{\begin{equation}}
\newcommand{\eeq}{\end{equation}}

\newcommand{\dd}{D\hspace{-.65em}/}

\def\op{operator}

\def\zm{zero mode}

\def\cn{condition}

\begin{document}
\begin{titlepage}
\begin{flushright}
FERMILAB-PUB-93/062-T
\end{flushright}
\vspace{0.5cm}
\begin{center}
{\large {\bf Twisting of N=1 SUSY Gauge Theories and
Heterotic Topological Theories}}\\
\vspace{1.5cm}
{\bf A.Johansen}
\footnote{e-mail:johansen@lnpi.spb.su, ajohansen@nbivax.nbi.dk}\\
\vspace{0.4cm}
{\em St.Petersburg Nuclear Physics Institute\\
Gatchina, St.Petersburg District, 188350 Russia}\\

\end{center}

\begin{abstract}
It is shown that $D=4$ $N=1$ SUSY Yang-Mills theory with an appropriate
supermultiplet of matter can be twisted on compact K\"ahler manifold.
The conditions of cancellation of anomalies of BRST charge are found.
The twisted theory has an appropriate BRST charge.
We find a non-trivial set of physical operators defined as classes
of the cohomology of this BRST \op .
We prove that the physical correlators are independent on
external K\"ahler metric up to a power of a ratio of two
Ray-Singer torsions for the Dolbeault cohomology complex
on a K\"ahler manifold.
The correlators of local physical \op s turn out to be independent of
anti-holomorphic coordinates defined with a complex structure on the
K\"ahler manifold.
However a dependence of the correlators on holomorphic coordinates
can still remain.
For a hyperk\"ahler metric the physical correlators turn out
to be independent of all coordinates of insertions of local
physical \op s.

\end{abstract}
\end{titlepage}

\newpage
\section{Introduction}
\setcounter{equation}{0}

It is well known that in supersymmetric (SUSY) theories certain correlators
do not depend on coordinates of inserted \op s due to the supersymmetric
Ward identities in the flat space-time \cite{vs}.
These are correlators of \op s which are the lowest components
of superfields of the same chirality.
In particular in the $D=4$ $N=1,2$ QCD
there are non-trivial correlators of this type which are proved to
be non-zero due to contributions of the instanton-like configuration in
the path integral \cite{affleck,vs2,amati}.
On the other hand
it has been discovered \cite{witten} that in the $N=2$ SUSY theories
the sector of theory which contains only the
\op s of the same chirality with correlators independent of
coordinates (from now on they will be refered
as the topological correlators) can be
mapped onto the space of local physical observables
of an appropriate topological theory.
Furthermore an independence of coordinates of the topological correlators
still holds in the topological theories even in the presence of
arbitrary non-trivial metric of space-time.
Usually these correlators do not depend also on coupling constants
and on external metric.
In topological theories one can also find non-local physical \op s
which can be represented as integrals over non-trivial cycles of
local operator-valued forms.
The independence of physical correlators on external metric and gauge coupling
constant turns out to be important
for description of various moduli spaces and smooth structures
of manifolds in two and four dimensions in terms of quantum field theory.

The procedure of translation of a SUSY model to a topological theory is
a twisting.
By twisting of supersymmetric models one can get a wide class
of topological theories.
In particular the most popular probably are topological Yang-Mills theories
\cite{witten},
topological $\sigma$ models \cite{witten2} and topological conformal
theories \cite{eguchi}.
In turn studying a ring of local observables in twisted theory
one can get much information about physical correlators in untwisted
SUSY theory \cite{phase,ring,vafa,pasq,cec2}.

In the present paper we focus on D=4 twisted SUSY Yang-Mills theories.
The topological Yang-Mills theory can be constructed through
a twisting of N=2 SUSY Yang-Mills theory \cite{ym,kr,galperin}.
Such a twisting procedure can be understood as a
modification of the energy-momentum
tensor of the theory by adding to it a derivative of an appropriate
non-anomalous (axial) current.
Such a modification of the energy-momentum tensor leads to a change of
dimensions and spins of the quantum fields.
As a consequence one of supergenerators becomes a scalar
one and can be interpreted as the BRST \op \ of the twisted theory.

Technically
for the case of N=2 SUSY the twisted version of the theory can be
obtained in the framework of the N=2 SUGRA$+$YM theory \cite{ym,kr,galperin}.
The supermultiplet of the supergravity fields contains vierbein field
$e_{\mu}^a$, gravitino
field $\psi_{\mu}^i$, $SU(2)$ valued vector field $V^i_{\mu j}$
(this $SU(2)$ symmetry generalizes the
$U(1)$ $R$-symmetry of $N=1$ SUSY for the case of $N=2$ SUSY) and other
(auxiliary) components ($\mu,\;\nu ...$ are world indices,
$a,\;b,..$ are indices in the tangent frame,
while $i$ and $j$ stand for
$SU(2)$ indices corresponding to the $SU(2)$ symmetry of the model).
The physical components of the Yang-Mills supermultiplet
are the gauge vector field $A_{\mu}$, gluino $\lambda^{\alpha}_i,\;
\bar{\lambda}^j_{\dot{\alpha}}$ and
complex scalar field $\phi$ (here $\alpha$ and $\dot{\alpha}$ are
the spinor indices).
The gauge group is assumed to be compact, finite dimensional.

This theory is invariant under simultaneous supergravity transformations
(which include localized SUSY and $SU(2)$ transformations)
of both supergravity and Yang-Mills multiplets.
To construct a topological theory
we are to consider the supergravity multiplet as a set of external fields.
In general all the symmetries generated by the $N=2$ supercharges
are broken in the presence of an external supergravity fields.
However for a special choice of the supergravitational fields
some of supersymmetries may still remain.
The prescription for such a choice is that one should put
all components of supergravity multiplet equal to zero except of
the veirbein field $e_{\mu}^a$ and $SU(2)$ vector component $V^i_{\mu j}$.
The auxiliary field $V^i_{\mu j}$ is coupled to a current that is used
for twisting of the Lagrangian of the SUSY theory on the curved manifold.
We are looking for a supercharge that survives as a symmetry of the theory.
It is clear that such a supercharge should not generate any
other components of the supergravity multiplet
in addition to the veirbein and the vector fields.
It is easy to see that this condition is equivalent to vanishing
of the variation of the gravitino component under such a supertransformation:
\beq
(\delta \psi_{\mu}^i)_{L,R}
=\; D_{\mu j}^i(\omega,V)\epsilon_{L,R}^j\; =
[\delta^i_j (\partial_{\mu} - \frac{i}{4} \omega_{\mu}^{ab}
\sigma_{ab}) \; \pm \; V_{\mu j}^i ]\epsilon_{L,R}^j =0,
\eeq
where the indices $L$ and $R$ stand for the left- and right-handed components
of spinor.
Here $\omega^{ab}_{\mu}$ is the usual spin-connection
\beq
\omega^{ab}_{\lambda}\; =\;
\frac{1}{2}
[e^{a\nu} e^{b\mu} e_{\lambda c} (\partial_{\nu} e^c_{\mu} -
\partial_{\mu}
e^c_{\nu}) - e^{a\nu} (\partial_{\lambda} e^b_{\nu}
- \partial_{\nu} e^b_{\lambda}) - e^{b\nu} (\partial_{\nu} e^a_{\lambda} -
\partial_{\lambda} e^a_{\nu})].
\eeq
It easy to see that the variation (1.1) can be
equal to zero for an appropriate choice of a constant spinor
$\epsilon$ and the
vector field $V_{\mu j}^i$ in any external veirbein field.
Thus the vector field is a functional
of veirbein field and depends on the choice of a spinor $\epsilon$.
The spinor $\epsilon$ is a parameter of global SUSY transformation which
survives in the twisted theory on the curved manifold.
Moreover this supercharge plays the role of the BRST charge
that allows to project out nonphysical (nonsuperinvariant) states
because the BRST charge is nilpotent and commutes with
the Hamiltonian of the theory.
The physical states correspond to the cohomology classes of the BRST charge.
The ring of BRST closed physical operators turns out to be
non-trivial and their correlators
are independent of external metric and gauge coupling constant.
However these correlators can depend on moduli of
a differential structure on the curved manifold.
Therefore
this theory gives a realization of the Donaldson map \cite{don} relating
a smooth structure of a 4-dimensional manifold to the topology of the
instanton moduli space in terms of physical correlators.
The physical correlators turn out to be Donaldson's invariants which
characterize smooth structures on 4D-manifolds.

This theory is also intriguing from the physical point of view.
In particular in the heterotic string theory the scattering
amplitudes of spacetime axions at zero momenta are found to be proportional to
the Donaldson invariants \cite{harvey}
in the form as they are represented in Witten's theory \cite{witten}.

In this paper we consider the case of $N=1$ SUSY gauge theories and their
twisted versions.
The motivation of this work is the following.
This theory also has a sector of chiral \op s whose correlators do not
depend on coordinates in the flat external metric \cite{vs,affleck,vs2,amati}.
The problem that we study is
to try to pick out the topological sector of the theory as space
of physical states selected by a BRST \op \ in a appropriately twisted theory.
We argue that the twisting procedure can be correctly done for
certain theories with non-anomalous $U(1)$ current on
compact, without boundaries, $D=4$ K\"ahler manifolds.
The condition of absence of anomaly of twisting $U(1)$ current is
shown to be equivalent to a certain condition to a gauge group representation
of the fields matter.
It comes from a condition of cancellation of a mixed anomaly
of gauge and gravitational fields \footnote{The
idea of a possibility of such a cancellation of anomaly
was suggested by A.Lossev.}.
We find a non-trivial
set of physical \op s defined as classes of the cohomology of the BRST \op .
In such a twisted theory physical correlators turn out to be
invariant under smooth deformations of external K\"ahler metrics up to
a power of a ratio of two Ray-Singer torsions for the Dolbeault cohomology
on K\"ahler manifold.
The correlators of the local observables turn out to be meromorphic
sections of a tensor power of a
holomorphic bundle over D=4 K\"ahler manifold $M$
with respect to the chosen complex structure and independent on
gauge coupling constant.
As a result the theory allows for a localization similar to Witten's theory
\cite{witten}.
As in the original SUSY theory the physical correlators
can be non-vanishing due to contributions of instanton (or anti-instanton)
classical configurations.
For the case of hyperk\"ahler manifolds the correlators of local
physical \op s turn out to be independent on coordinates of the \op .

We will henceforth generically name such (non-anomalous)
twisted theories on K\"ahler manifold heterotic topological theories.
A heterotic topological theory for a particular
case when the fields of matter belong to an adjoint representation of
the gauge group turns out to be equivalent to Witten's theory.
However it is different from Witten's theory when the fields of
matter transform as non-adjoint representation.
In particular in a generic situation
the correlators in heterotic topological theory depend on complex structure.
Therefore we can hope that the correlators in the heterotic theory can
give additional information about inavariants
of smooth and complex structures of D=4 K\"ahler manifolds.

It is also interesting that
a heterotic theory seems to give a possibilty for geometrical
interpretation of zero modes of fields of matter in non-adjoint
representation of the gauge group.

The paper is organized as follows.
In sect.2 we discuss the \cn s under which there is an unbroken supersymmetry
on a curved manifold and describe the twisting procedure in terms of
the fields of SUSY theory at the classical level.
In sect.3 the physical observables are defined and the Ward identities for
physical correlators are found.
The model is reformulated in terms of sections of holomorphic and
anti-holomorphic vector bundles on the 4-manifold.
In sect.4 and 5 we analyse the semiclassical representation of
the physical correlators and deduce the \cn s of cancellation of
anomalies of BRST charge.
The structure of the physical correlators is discussed.
In Conclusions the results of this paper are summarized and some
possibilities for further investigations are discussed.

\section{N=1 SUSY Yang-Mills theory in an external gravitational field}
\setcounter{equation}{0}

We start with the N=1 Yang-Mills theory coupled to external supergravity
multiplet in the framework of "new minimal supergravity" \cite{west}
(see also \cite{ferrara}).
The supergravity multiplet contains veirbein field $e_{\mu}^a$,
gravitino field $\psi^{\alpha}, \;\bar{\psi}_{\dot{\alpha}}$,
the $U(1)$ vector field $V_{\mu}$ and auxiliary fields.
The physical components of Yang-Mills superfield are the gauge field
$A_{\mu}$ and gaugino $\lambda^{\alpha},\; \bar{\lambda}_{\dot{\alpha}}$.
This theory is invariant under simultaneous
local SUSY transformations for super Yang-Mills fields
and external supergravity multiplet.

For definiteness the external metrics is assumed to be
euclidean with the signature $(+,+,+,+)$
(that means that one should consider
$\lambda^{\alpha}$ and $\bar{\lambda}_{\dot{\alpha}}$ as independent fields).
We are just looking now for a generator of
SUSY transformations that can still correspond to a symmetry of the model
if we reduce the supergravity multiplet to the usual metric
and an external vector field built out of the veirbein field (these
fields of supergravity multiplet are considered as external ones).
If such a generator does exist it should correspond to a global
supertransformation which does not change the external supergravity fields.
It can be easily checked that only gravitino field can be induced
under infinitesimal supertransformations.
Such a variation of the gravitino fields reads as follows
\beq
(\delta \psi_{\mu})_{L,R}
=\; \nabla_{\mu}(\omega,V)\epsilon_{L,R},
\eeq
where
\beq
\nabla_{\mu}(\omega ,V) = (\partial_{\mu} - \frac{i}{4} \omega_{\mu}^{ab}
\sigma_{ab} \; \pm \; V_{\mu})\epsilon_{L,R}.
\eeq
Here the latters $\mu ,\; \nu ...$
stand for world indices while $a,\; b,\;
c...$ correspond to the Lorentz indices in the tangent frame.

{}From eq.(2.1) one can see that the residual global supersymmetry still
holds if the matrix in the twisted covariant derivative for spinor
representation has the same constant eigenvector $\epsilon$
for all $\mu$, i.e.
\beq
(\frac{i}{4} \omega_{\mu}^{ab} \sigma_{ab} \;+\;
V_{\mu} \gamma_5)\;\epsilon\;=\;0.
\eeq
It is clear that this equation can not be satisfied for a generic
external metric (the difference with the twisting of $N=2$ case
is that here we have a smaller number of parameters in the
twisting vector field).
However for some special manifolds it can have solutions.
To classify all such situations let us recall that
the euclidean Lorentz group
$SO(4)$ is isomorphic to $SU(2)\times SU(2)$ group that
acts in the spinor space as $SU(2)$ for both chiralities.
That means that in the presence of a non-flat external metric the holonomy
group is generically $SU(2)\times SU(2)$.
However for special manifolds it can be reduced.
There are two cases (see, e.g. \cite{gsw}) of reduction of holonomy
group which are interesting for us: $SU(2)\times SU(2) \to U(2)$
and $SU(2)\times SU(2) \to SU(2)$.
The first case corresponds to
the K\"ahler metrics while the second defines the hyperk\"ahler one.
It is worth emphasizing that on K\"ahler manifolds, dotted and undotted
spinor indices are essentially different.
A K\"ahler structure distinguishes one of chiralities.

Let us consider, for example, the case when the $SU(2)$
group acting on the right-handed sector is reduced.
Notice that in the case of hyperk\"ahler metric no vector field $V_{\mu}$ is
necessary to make a twisting because the right-handed projection of
the matrix $ \omega_{\mu}^{ab} \sigma_{ab}$ vanishes (up to a pure gauge
transformation of the Lorentz group).
Moreover there are two global supersymmetries
with right-handed generators which do not affect the external metric.
In turn for the K\"ahler manifold the matrix
$ \omega_{\mu}^{ab} \sigma_{ab}$ in the rigt-handed sector can be taken
locally proportional to a fixed constant $\sigma$ matrix, for example,
to the $\sigma_3$ matrix in an open region on the 4-manifold.
Then it is easy to see that eq.(2.3) can be satisfied
by an appropriate choice of vector field $V_m$ and
a spinor $\epsilon$.
Therefore the supercharge
corresponding to $\epsilon$ generates a supersymmetry of the system.

The lagrangian of the twisted SUSY Yang-Mills theory in the external
gravitational field reads as
\beq
L = \sqrt{g}
Tr\; [\frac{1}{4}
FF\; +\; \bar{\lambda} i\dd \;\lambda\; +\;\frac{1}{2}
D^2]
\eeq
where $\dd = e_a^{\nu} \dd_{\nu} \gamma_a ,$ $D_{\nu} =\nabla_{\nu}
(\omega ,V)- iA_{\nu} ;$
$\bar{\lambda}$ and $\lambda$ stand for the right and left-handed Weyl
spinor correspondingly, $F_{\mu \nu}$ is the strength tensor for the
Yang-Mills field $A_{\mu}$ and $D$ is
an auxiliary scalar field which is necessary for SUSY algebra to be
closed.
We assume that the spin-connection $\omega^{ab}_{\mu}$
in the sector of right-handed fermions is proportional to
the generator of the $U(1)$ group
or vanishes.
The global transformation with a right-handed
supergenerator which corresponds to the constant spinor
$\bar{\epsilon}$ obeying eq.(2.3) is as follows
(we use the Weyl notations)
\beq
\delta A_{\mu}\;=\;(\bar{\epsilon} \sigma_a \lambda) e^a_{\mu} \; ,
\;\;\delta \lambda \;=\;0\;,
\eeq
$$\delta\bar{\lambda}\;=\; D \bar{\epsilon} -\frac{1}{2}\;
F_{\mu \nu}\bar{\epsilon}
\sigma^{ab} \;e^{\mu}_{\nu} e^{\nu}_b,\;\;
\delta D\;=\; -\bar{\epsilon} i\dd \lambda .$$
One can show by a direct calculation that the Lagrangian (2.4)
is invariant under the transformation (2.5).
Actually the variation of the lagrangian with an arbitrary
constant spinor $\epsilon$ is proportional to the same
expression that appears for the variation of gravitino field (eq.(2.1))
$$\hat{F}^{\mu \nu}\; \bar{\epsilon}
\;[\frac{i}{4} \omega_{\mu}^{ab} \sigma_{ab} \;+\; V_{\mu}]
\;e^c_n \sigma_c \lambda\;$$
which is zero precisely for the $\bar{\epsilon}$ obeying eq.(2.3).
Here $\hat{F}^{\mu \nu}$ stands for self-dual
part of the Yang-Mills strength.

It can be shown that the \op \ $Q$ which corresponds to
the supergenerator associated with the spinor $\bar{\epsilon}$
is nilpotent, i.e.
\beq
Q^2\;=\;0.
\eeq
Another important fact is that this lagrangian is $Q$-exact.
By a direct calculation one can check that
\beq
L\;=\;
\frac{1}{2} \left\{ Q,\; \bar{\lambda} \left(
D\;+\;\frac{1}{2}\sigma^{\mu \nu} F_{\mu \nu}\right) \bar{\eta}\right\} .
\eeq
Here the constant spinor $\bar{\eta}$ is linearly independent of
$\bar{\epsilon}$ and obeys eq.(2.3) with a change $V_{\mu} \to
-V_{\mu} .$
Its normalization is defined by the condition
$\bar{\epsilon}_{\dot{\alpha}} \bar{\eta}_{\dot{\beta}}
\;\epsilon^{\dot{\alpha} \dot{\beta}}\; =\;1$
(here $\epsilon^{\dot{\alpha}\dot{\beta}}$ is the conventional metric
in the usual spinor bundle).
Eq.(2.7) for the Lagrangian is equivalent to eq.(2.4)
up to the terms that vanish for the spinor $\epsilon$ obeying eq.(2.3) and
a term which is proportional to a topological charge of the gauge field.

A similar procedure can be also done for a chiral supermultiplet of matter
in any representation of the gauge group and coupled to external
supergravity.
It is important however that the axial current of fermions which belong to the
supermultiplet of matter
is coupled to the twisting vector field $V_{\mu}$ with
an opposite charge as compared to the charge of the gaugino current.
This fact will allow us to cancel an anomaly of a total axial current
which is used for twisting.
An absence of anomaly for this current is important when we want to
interprete the twisted theory as a topological one.

A chiral supermultiplet of matter in a representation $R$ of the gauge group
contains a complex scalar field $\phi$ and
a Weyl fermion $\psi_{\alpha}$ and also an auxiliary complex scalar field $F$
($\bar{\phi} ,$ $\bar{\psi}_{\dot{\alpha}}$ and $\bar{F}$ are the
complex conjugated fields, respectively).
The lagrangian of matter coupled to external metric $g_{\mu\nu}$ and
the vector field $V_{\mu}$ and to the Yang-Mills supermultiplet reads
\beq
L_{matter} = \sqrt{g} (- D_{\mu}\bar{\phi} D^{\mu} \phi +
\bar{\psi} i \dd \psi + \bar{F} F -
\bar{\phi} D \phi -2 \bar{\phi} \lambda \psi + \bar{\psi}\bar{\lambda} \phi ).
\eeq
It is invariant under the transformations (2.5) and
the transformations of the fields of matter which read as follows
\beq
\delta \phi = 0 ,\;\;\;
\delta \psi = i \sigma_{\mu} \bar{\epsilon} D^{\mu} \phi ,\;\;\;
\delta F = -\bar{\epsilon} i\dd \psi - \bar{\lambda} \bar{\epsilon}
\phi ,
\eeq
$$\delta \bar{\phi} = \bar{\epsilon} \bar{\psi},\;\;\;
\delta \bar{\psi} = \bar{F} \bar{\epsilon} ,\;\;\; \delta \bar{F} =0.$$
Here $D_{\mu}$ is a covariant derivative with respect to the gauge field
when acting to a scalar field, while it contains the twisted
spin-connection when acting to a spinor field.
The lagrangian of matter (2.8) is $Q$-exact
$$L_{matter} =\{ Q, \bar{\eta}(- \bar{\psi} F + \bar{\phi} i\dd \psi
+ \bar{\phi} \bar{\lambda} \phi ) \} .$$
The total supercharge $Q$ is nilpotent.

It is clear now that any physical correlators with insertions
of $Q$-exact \op s
vanish due to $Q$-invariance of the action.
In turn it is easy to see that the left-handed gluino field $\lambda$
is $Q$-invariant and is not $Q$-exact.
Therefore there is a natural candidate for the gauge-invariant
non-trivial local observable given by the \op \
\beq
Tr\;\lambda \lambda,
\eeq
where the scalar product is defined with the conventional
constant antisymmetric metric $\epsilon_{\alpha \beta} .$
However this \op \ is not quite appropriate for our purpose.
The point is that in general it is not a scalar \op .
Actually while in the untwisted theory
the \op \ $Tr\;\lambda \lambda$ is scalar with respect to
the usual spin-connection, in a twisted version of the model
it aquires a non-trivial axial charge
due to coupling of $\lambda$ to the external vector field $V_{\mu}$
which enters the definition of the paralell transport on the spinor
bundle on $M$ and hence it is not a scalar.
Actually one can check by a direct calculation
that
\beq
(\partial_{\nu} \;+\;2 V_{\nu})\;(Tr\lambda \lambda)
\;\sigma^a e^{\nu}_a \bar{\epsilon}
\;=\; -2i\;\left\{ Q, Tr\left[ \left( D\;+\;
\frac{1}{2}\sigma^{\mu \nu} F_{\mu \nu} \right) \lambda \right]
\right\}.
\eeq
This is not quite that equation which would allow us
to prove an independence of the correlators
of the \op s $Tr\;\lambda \lambda$ on coordinates.
We have to define an \op \ the usual derivative of which is given by a
$Q$-exact \op .
Thus one should redefine fields and metric on the spinor bundle
in order to eliminate the vector field $V_{\nu}$ from the left hand
side of this.
We consider this problem in the next section.

In the sector of fields of matter the local observables
are given by gauge invariant combinations of the scalar fields, e.g.
$<\phi \phi>$ (here $<...>$ stands for a scalar product for a representation
$R$ of the gauge group).
{}From eqs.(2.9) we can see that some linear combinations of usual
derivatives in coordinates of $<\phi \phi>$ are $Q$-exact.
Therefore the correlators with insertions of these derivatives of
local operators vanish.
This is a generalization of supersymmetric Ward identities
to the case of curved manifold.
Actually the set of local scalar \op s depends on the gauge group and on
a representation of the multiplet of matter and can include ``mesons'' and
``baryons'' \cite{seiberg}.
These local \op s correspond to flat directions in the classical
moduli space of vacua in SUSY theory.

It is possible to define also non-local $Q$-invariant operators
as integrals of gauge invariant \op s
integrated with some appropriate functions, but we postpone
a description of such \op s for the next sections.

Notice also that if the structure of the sector of the fields of matter allows
for N=2 SUSY of the whole
untwisted theory in flat space then this SUSY theory can be of course
twisted in a different way due to $SU(2)$ global symmetry of N=2 SUSY.
Such a twisting can be done on any 4D manifold.
However we shall not consider this case in the present paper.

\section{Observables and Topological Correlators}
\setcounter{equation}{0}

To understand better eq.(2.11) it is useful to define a complex structure
$J^{\mu}_{\nu}$ on the manifold $M$ in terms of the (chosen locally to
be constant) spinors $\bar{\epsilon}$ and $\bar{\eta}$:
\beq
J^{\mu}_{\nu} = (\bar{\eta} \sigma^{ab}\bar{\epsilon}) e^{\mu}_a\; e_{\nu b}.
\eeq
Indeed it is easy to check that
\beq
D_{\lambda} J^{\mu}_{\nu} =0\eeq
where $D_{\lambda}$ is the nontwisted covariant derivative,
and hence $J^{\mu \nu}$ is a closed two-form
\beq
dJ =0.
\eeq
Moreover $J^{\mu \nu}$ is anti-selfdual so that the two-form $J$ is harmonic
\beq
(^* d^* d + d ^* d^*)J = 0,
\eeq
and
\beq
(J^2)^{\mu}_{\nu} = - \delta ^{\mu}_{\nu}.
\eeq
The complex structure allows to define holomorphic and anti-holomorphic
coordinates
\beq
J^m_n\;z^n =i z^m,\;\;J^{\bar{m}}_{\bar{n}}
\;\bar{z}^{\bar{n}} =-i\bar{z}^{\bar{m}},
\eeq
where $m,\; n$ and $\bar{m},\; \bar{n}$ takes values 1,2 and $\bar{1},\;
\bar{2}$ respectively.
In the well adapted frame
(see, e.g., \cite{gsw}) the complex structure has the simplest
form
\beq
J^m_n = i\delta^m_n,\;\;
J^{\bar{m}}_{\bar{n}} = -i\delta^{\bar{m}}_{\bar{n}},
\eeq
$$J^{m n} =J^{\bar{m} \bar{n}} =0,$$
and hence
\beq
J^{m \bar{n}} = ig^{m \bar{n}},
\eeq
while the metric
has only $g^{m \bar{n}}$ non-vanishing components.
Now we can rewrite eq.(2.11) in an equvalent form in terms of
$z_n$ and $\bar{z}_{\bar{n}}$ coordiantes.
To this aim let us consider the scalar product of both sides
of this equation with a spinor $\bar{\eta} \sigma_{\bar{k}}$.
Then one can get the following equation
\beq
e^{-1} \partial_{\bar{k}} (e Tr\lambda \lambda) =
2\;\left\{ Q, Tr\left[ \bar{\eta}\sigma_{\bar{k}}
\left( D\;+\;
\frac{1}{2}\sigma^{mn} F_{mn} \right) \lambda \right]
\right\}.
\eeq
Here $e= {\rm det} e_m^a$ while $e_m^a$ stands for components of
the veirbein field with both holomorphic world index $m$ and Lorentz
index $a$ (we use the tangent frame where the real veirbein field
$e^a_{\mu}$ is block-diagonal; from now on we shall use the letters
$a,\; b...$ for holomorphic indices and $\bar{a},...$ for anti-holomorphic
ones).
Thus we find that the derivative of the \op \
\beq
O_0 = \exp(-2\int^z V_{\bar{n}}
dz^{\bar{n}}) {\rm Tr} \lambda\lambda = e{\rm Tr}\lambda\lambda
\eeq
in anti-holomorphic coordinates is $Q$-exact.
Here we used that the twisting vector field $V_{\mu}$ locally is a total
derivative: $V_n = (1/2)\partial_n \log \bar{e}$ and
$V_{\bar{n}} = -(1/2)\partial_{\bar{n}} \log e .$
However globally on $M$ the exponential in the definition of the \op \ $O_0$
can depend on the contour of integration, and therefore
in general this \op \ is not well defined globally.
As it will be discussed in the next section
there could be still an interesting interpretation of the \op \ $O_0 .$
There is no such a problem for the local \op s constructed of the
field $\phi$ of matter since it has no axial charge and, hence, these
\op s are well defined globally.

Thus if the BRST symmetry does not have any anomalies at the quantum
level the correlators of local \op s are independent of
anti-holomorphic coordinates by the standard arguments.

Notice that for the case of hyperk\"ahler manifold there is also another
equation due to the second BRST charge (the twisting field $V_{\mu}$
can be gauged out in this case).
Using the second BRST charge one can show that the correlator of
these physical \op s does not depend on all coordinates.

Now we want to understand if there is any dependence of physical correlators
on external K\"ahler metric.
We observe that the metric enters both the definition of the $Q$
transformation and the action.
Furthermore the variation of the action in the metric is not $Q$-exact
since the definition of the \op \ $Q$ depends on the metric.
To overcome this difficulty it is necessary to redefine
the quantum fields in the theory and to reformulate the definition of
the Lagrangian,
observables  and the BRST charge in terms of sections of holomorphic
and anti-holomorphic vector bundles over the manifold $M .$

Let us consider first the sector of the gauge multiplet.
Let us introduce the following fields:
\beq
\chi_n = \bar{\epsilon} \sigma_n \lambda,\;\;
\bar{\lambda}_{\bar{m} \bar{n}} =
e^{\bar{m}}_{\bar{a}} e^{\bar{n}}_{\bar{b}}\;
(\bar{\eta} \sigma_{\bar{a} \bar{b}} \bar{\eta})
\;(\bar{\epsilon} \bar{\lambda})
,\;\; \bar{\lambda}= \bar{\eta}\bar{\lambda},
\eeq
$$D' = D + i J^{\bar{m} n} F_{\bar{m} n} ,$$
where $F_{m\bar{n}}$ is the strength tensor components of the
Yang-Mills field.
Now the right-handed spinor $ \bar{\lambda}_{\dot{\alpha}}$
is splited into zero-form $\bar{\lambda}$ and the anti-holomorphic
two-form $\bar{\lambda}_{\bar{m} \bar{n}}$.
The left-handed spinor $\lambda_{\alpha}$ becomes
a section $\chi_n$ of the holomorphic vector bundle over $M.$
The field $D$ is still an auxiliary zero-form
but is shifted by the strenght tensor of the gauge field.

The BRST transformation rules in terms of these forms
do not now depend on external metric:
\beq
\delta A_n = \chi_n,\;\;\delta A_{\bar{n}} =0,\;\;
\delta \chi_n =0,
\eeq
$$\delta \bar{\lambda} = - D',\;\;
\delta \bar{\lambda}_{\bar{m} \bar{n}} =
2i F_{\bar{m} \bar{n}},\;\;
\delta D' = 0.$$
One can see that all the fields are combined into three different
supermultiplets: ($A_n ,\; \chi_n$), ($\bar{\lambda},\; D'$) and
($\bar{\lambda}_{\bar{m} \bar{n}},\; F_{\bar{m} \bar{n}}$).
Fixing the ghost number of the BRST charge to be 1 we have
the following dimensions ($d$) and
ghost numbers ($G$) of the fields:
\beq
(d,G)(A_n) = (d,G)(A_{\bar{n}}) = (1,0), \;\;
(d,G)(\chi_n) = (1,1)\;\;
\eeq
$$(d,G)(\bar{\lambda}) = (d,G)(\bar{\lambda}_{\bar{m} \bar{n}})=
(2,-1) ,\;\;
(d,G)(D') = (2,0).$$
The conservation of the ghost number is broken by an anomaly.

The Lagrangian for the gauge multiplet reads now as follows
\beq
L = \sqrt{g} Tr [ F^{\bar{m}\bar{n}} F_{\bar{m}\bar{n}}
+i \bar{\lambda}^{mn} \nabla_m \chi_n +
\eeq
$$+ \frac{1}{2} D'^2 + J^{\bar{m} n} (iD' F_{\bar{m} n}
+\bar{\lambda} \nabla_{\bar{m}} \chi_n)] =$$
$$= \sqrt{g}
\{ Q, Tr[-\bar{\lambda}  J^{\bar{m} n} F_{\bar{m} n} -
\frac{1}{2} D' \bar{\lambda} -
\frac{i}{2} \bar{\lambda}_{\bar{m}\bar{n}}
F^{\bar{m}\bar{n}}] \} ,$$
where $g$ is the determinant of the metric.
It is easy to see that a variation of this Lagrangian in
external metric is $Q$-exact.

Let us consider now the fields of matter.
After an appropriate redefinition of the fields of matter
we get two scalar fields $\phi$ and $\bar{\phi}$, a scalar fermion
$\bar{\psi}$,
the fermionic fields $\psi_{\bar{m}}$ and $\bar{\psi}^{\bar{m}\bar{n}}$
which are $(0,1)$ and $(2,0)$ forms respectively, and
the auxiliary $(0,2)$ and $(2,0)$ forms $N_{\bar{m}\bar{n}}$ and
$\bar{N}^{\bar{m}\bar{n}} .$
The BRST transformations (for matter coupled to the gauge supermultiplet)
read now as follows
\beq
\delta \phi =0,\;\;\; \delta \psi_{\bar{m}} = D_{\bar{m}}
\phi ,\;\;\; \delta N_{\bar{m}\bar{n}}= D_{\bar{m}}
\psi_{\bar{n}} - D_{\bar{n}}\psi_{\bar{m}} +\frac{1}{2}
\bar{\lambda}_{\bar{m}\bar{n}} \phi ,
\eeq
$$\delta\bar{\phi} =\bar{\psi} ,\;\;\;
\delta \bar{\psi} =0 ,\;\;\;
\delta \bar{\psi}^{\bar{m}\bar{n}} = -2 \bar{N}^{\bar{m}\bar{n}} ,\;\;\;
\delta \bar{N}^{\bar{m}\bar{n}} =0 .$$
The derivative $D_{\bar{m}}$ and $D^{\bar{m}}$ are covariant with respect
to both external metric and the gauge field.
The lagrangian of matter has the following form
\beq
L_{matter}=
\sqrt{g} (\bar{\phi} D^{\bar{m}}D_{\bar{m}}
\phi + \bar{\psi}^{\bar{m}\bar{n}}
D_{\bar{m}} \psi_{\bar{n}} + \bar{\psi}
D^{\bar{m}} \psi_{\bar{m}} + N_{\bar{m}\bar{n}}
\bar{N}^{\bar{m}\bar{n}} -
\eeq
$$- i\bar{\phi} \chi_n \psi^n - \bar{\psi} \bar{\lambda} \phi +
\frac{1}{4} \bar{\psi}_{mn}\bar{\lambda}^{mn} \phi  +
\bar{\phi} D \phi) .$$
This lagrangian is $Q$-exact
\beq
L_{matter} = \{ Q, -\frac{1}{2} \bar{\psi}^{\bar{m}\bar{n}} N_{\bar{m}\bar{n}}
+ \bar{\phi} D^{\bar{m}} \psi_{\bar{m}} - \bar{\phi} \bar{\lambda} \phi \} ,
\eeq
while the total BRST charge is nilpotent.

Thus the total lagrangian of the gauge multiplet + matter is BRST exact
and formally allows for a localization because its variations in the
external metric and in the gauge coupling constant are BRST exact.
The fields of the multiplet of matter have the following dimensions ($d$) and
ghost numbers ($G$)
\beq
(d,G)(\phi) = (0,2),\;\;\;(d,G)(\bar{\phi})=(2,-2),\;\;\;
(d,G)(\psi_{\bar{m}})=(1,1),
\eeq
$$(d,G)(\bar{\psi})=(d,G)(\bar{\psi}_{mn})=(2,-1),\;\;\;
(d,G)(N_{\bar{m}\bar{n}})=(2,0), \;\;\; (d,G)(\bar{N}_{mn})=(2,0).$$

The local observable for the sector of the gauge multiplet
becomes a (2,0) form (of dimension 2)
\beq
O^{(0)}_{mn} = Tr \chi_m \chi_n,
\eeq
instead of $O_{0}$ (see eq.(3.10)), which is related to $O^{(0)}_{mn}$
by the following equation
\beq
O_{0} = e Tr \chi_m \chi_n \; (\bar{\epsilon} \sigma_{ab}
\bar{\epsilon})\; e^{a m} e^{b n}.
\eeq
It is to be noticed that the situation is different from the
ordinary topological theories where the local observables are zero-forms;
non-zero forms should usually be integrated over closed cycles
to get non-local observables (in the case of highest forms one gets
moduli of the topological theory).
The difference here is due to the splitting of four coordinates
into holomorphic and anti-holomorphic ones, so that the (2,0) form
is effectively a scalar with respect to anti-holomorphic derivatives.
We have
\beq
\partial_{\bar{k}} Tr \chi_m \chi_n
= \{ Q,...\} .
\eeq
It follows from this equation that the physical correlators with insertions
of this operator are holomorphic with respect to its coordinate.
However we will see that in a semiclassical representation such a
correlator is given by an integral over the moduli space of instanton.
In turn such an integration could induce singularities in the correlator.
Notice also that as we saw in the SUSY version of the theory the analog of
the \op \ $O^{(0)}$ in SUSY theory is not in general globally defined.
A manifestation of that here is that the \op \ $O^{(0)}_{mn}$ is a section of
a holomorphic bundle over $M .$
The correlators with insertions of such an \op \ can have a nontrivial
monodromy on $M .$
It is tempting to try to interprete these correlators as a generalization of
conformal blocks of 2D conformal theories (which usually have
a non-trivial monodromy on 2D manifold; see, e.g. \cite{ginsparg})
to the 4D case.
We can also integrate such an \op \ with an appropriate
closed (0,2) form: then we get a well defined non-local \op .

One can construct also the non-local observables using the method of discent
equations \cite{witten}.
We have
\beq
\partial_{\bar{k}} O^{(0)}_{mn} =
\{ Q, H^{(1)}_{mn, \bar{k}}\},
\eeq
$$\partial_{[\bar{p}}  O^{(1)}_{mn, \bar{k}]} =
\{ Q , H^{(2)}_{mn, \bar{k}\bar{p}}\}$$
where
\beq
H^{(1)}_{mn, \bar{k}}=
Tr (F_{\bar{k} m} \chi_n -F_{\bar{k} n} \chi_m ) ,
\eeq
$$H^{(2)}_{mn, \bar{k}\bar{p}} =
\frac{1}{2} Tr (F_{\bar{k} m} F_{\bar{p} n}-
F_{\bar{k} n} F_{\bar{p} m} + F_{\bar{p}\bar{k}}
F_{m n}) +
\frac{i}{4} (\partial_m Tr \bar{\lambda}_{\bar{p} \bar{k}}
\chi_n - \partial_n Tr \bar{\lambda}_{\bar{p} \bar{k}}
\chi_m).$$
The \op \ $H^{(2)}_{mn, \bar{k}\bar{p}}$ is obviously
the density of the topological charge of the gauge field up to
an exact form.
These relations allows us to construct the following
non-local observables
\beq
O^{(1)} =\int \bar{\omega} H^{(1)},
\;\;
O^{(2)} = \int  H^{(2)},
\eeq
where $\omega$ is a closed (0,1) form.
Here we used that the forms $H^{(1)}$ and
$H^{(2)}$ are closed up to BRST-exact \op s according to eqs.(3.22) and (3.23).

The local \op s in the sector of matter are given by the same
gauge invariant functions of the (dimensionless) scalar field $\phi$ as in the
supersymmetric version of the theory (because this scalar field has zero
axial charge) which correspond to flat directions of the classical
moduli space of vacua \cite{seiberg}.
It is also possible to construct non-local operators.
In this paper we focus on the BRST-invariant \op \
\beq
O_{matter} = \sum_{IJ} a_{IJ}
\int E \wedge <\psi^I \wedge \psi^J + N^I \phi^J> ,
\eeq
where $<...>$ stands for a gauge invariant pairing of the fields
(its definition depends on a representation of the gauge group for the
fields of matter),
the indices $I$ and $J$ stand for different irreducible
multiplets of matter,
and $E$ is a holomorphic (2,0) form on $M .$
$a_{IJ}$ is a constant matrix
(a choice of it depends on a represenation of the fields of matter).
For example for the case when the multiplet of matter is in an adjoint
representation of the gauge group the matrix $a_{IJ}$ is just 1.
For the case of the gauge group $SU(2)$ and the fields of matter in two copies
of a spinor representation $a_{IJ}$ is antisymmetric.

It is worth noticing that we could use the vector field $-V_{\mu}$ for a
twisting of the theory.
Such a modification of the model corresponds to a change $\epsilon ,\eta
\to \eta ,\epsilon .$
The local \op s in this mirror model
are antiholomorphic up to BRST exact \op s (for example, $\partial_n \phi =
\{Q , ...\}$) while
their correlators \op s are anti-meromorphic.

\section{Instanton measure and Ray-Singer torsion}
\setcounter{equation}{0}

In this section we consider the \cn \ of cancellation of BRST anomaly
and discuss the structure physical correlators in semiclassical
representation.

Let us consider a calculation of a physical correlator.
First we observe that the term $(1/2)TrD'^2$ in the Lagrangian
for the sector of gauge multiplet
is BRST exact and can be (at least formally) taken out from the
Lagrangian without any change of the correlator.
After this modification one can integrate out over the field $D'$
that leads to the following constraint
\beq
J^{m\bar{n}} F_{m\bar{n}} =0.
\eeq
It is easy to see that this \cn \ is necessary for the anti-self-duality
of the Yang-Mills field.
To show that it is convenient to use the idenity
$\epsilon_{m \bar{n} k \bar{l}} =  J_{m \bar{n}} J_{ k \bar{l}}
-J_{m \bar{l}} J_{ k \bar{n}}$.
Then one can check that
\beq
\epsilon_{\bar{n} km \bar{l}} F^{m \bar{l}}
= - F_{\bar{n} k} + J_{\bar{n} k} (J^{\bar{p} q} F_{\bar{p} q}),
\eeq
$$\epsilon_{\bar{n} km \bar{l}} F^{km} = 2 F_{\bar{n}\bar{l}},
\;\;\; \epsilon_{\bar{n} km \bar{l}} F^{\bar{n}\bar{l}} = 2 F_{km}.$$
In turn the anti-self-duality \cn \ means that
\beq
\epsilon_{\bar{n} km \bar{l}} F^{m \bar{l}} = - F_{\bar{n} k} ,\;\;\;
\epsilon_{\bar{n} km \bar{l}} F^{km} = -2 F_{\bar{n}\bar{l}},\;\;\;
\epsilon_{\bar{n} km \bar{l}} F^{\bar{n}\bar{l}} = -2 F_{km}.
\eeq
{}From these equations it is easy to see that the anti-self-duality \cn \
is equivalent to the following equations:
\beq
J^{\bar{p} q} F_{\bar{p} q} =0,\;\; F^{km} = F^{\bar{n}\bar{l}} =0.
\eeq
Eq.(4.1) coincides with the first of the anti-self-duality \cn s.
The other \cn s in eq.(4.4) appear in the limit of weak gauge coupling constant
because in this case the functional integral is dominated by
the fields with $F^{km} = F^{\bar{n}\bar{l}} =0$
which correspond to the minimum of the action.
In turn we can consider this limit since the action of the theory is
BRST exact and, hence,
the correlators are independent of the value of the coupling constant.
Therefore the theory is semiclassical similar to Witten's theory and
the physical correlators can be computed
semiclassically in the presence of anti-instanton field.
In the presence of anti-instanton field some of fields of the model have
zero modes which should be substituted into the preexponential factor
in the path integral for correlator (directly or using Yukawa couplings).
Moreover one should integrate over quadratic fluctuations
near the anti-instanton field.
Notice that the classical action for anti-instanton equals to zero.

Let us analyse the \zm s.
Actually it is easy to see from eqs.(4.4) that the variation of the
self-duality equations in the vector field $A_n$ (fixing gauge of the variation
of the gauge field to be $D^{\mu} \delta A_{\mu} =0$) gives the following
equations
\beq
D_{[m } \delta A_{n]} = 0,
\;\; J^{\bar{m} n} D_{\bar{m}} \delta A_{n} = 0 ,
\eeq
\beq
D_{\bar{[m}} A_{\bar{n]}} =0,\;\;\;
J^{\bar{m} m}D_m \delta A_{\bar{m}} .
\eeq
We shall see that
up to gauge transformations these equations determine the zero modes
of the gauge field in the presence of anti-instanton field.

The equations of motion for the field $\chi_n$ read as follows
\beq
D_{[m } \chi_{n]} = 0,\;\;
J^{\bar{m} n} D_{\bar{m}}\chi_{n} = 0.
\eeq
Due to similarity of these equations to eqs.(4.5) we see that the
\zm s of the fermionic field $\chi_n$ and $A_n$ coincide.
Therefore the \zm s of $\chi_n$ correspond to a half of tangent vectors to the
moduli space ${\cal M}$ of anti-instanton because there is no superpartners
to $A_{\bar{n}} .$
As we shall see there is a natural complex structure on ${\cal M}$
and the \zm s of $\chi_n$ correspond to holomorphic tangent
vectors on ${\cal M} .$

For a compact manifold $M$
the moduli space ${\cal M}$ of anti-instanton is a manifold of dimension
\cite{freed}
\beq
d= p_1(G) -\frac{1}{2} {\rm dim} G (\chi -\tau) ,
\eeq
where $p_1(G)$ is the first Pontrjagin class of the adjoint bundle over $M,$
$G$ stands for the gauge group, $\chi$ is the Euler characteristic of
$M,$ and $\tau$ is the signature of $M .$
In this paper we focus on the case of generic irreducible anti-instanton field
and assume that the dimension of the moduli space ${\cal M}$ is given by
$p_1(G) .$
For simplicity we consider the case of the gauge group $SU(2)$
and an anti-instanton field with the Pontrjagin class
$p_1(SU(2)) = 8$ (the topological charge equals to -1).
In this case for a generic K\"ahler manifold
there are 8 zero modes for the gauge field and 4 zero modes for
the fermionic field $\chi_{n}$ and no \zm s for other fields (see, e.g.
cite{abc}).

Notice that the fermionic field $\chi_m$ has 4 zero modes
but not 8 despite the similarity of eqs.(4.5) and (4.7).
The point is that for the gluonic field we should consider the pairs
$(\delta A_m ,\delta \bar{A}_{\bar{m}})$
while the fermionic field has only components with holomorphic index.
It is easy to show that if $(\delta A_m ,\delta \bar{A}_{\bar{m}})$
is a wave function for a gluonic zero mode then
$(i\delta A_m ,-i\delta \bar{A}_{\bar{m}})$ also corresponds to a zero mode
because the latter can be rewritten as $J^{\mu}_{\nu} \delta A_{\mu}$
which satisfies the same equation as $\delta A_{\mu}$ since the tensor
$J^{\mu}_{\nu}$ is covariantly constant.

Let us consider now the sector of the fields of matter.
In the semiclassical limit the path integral is dominated
by the contributions from solutions to
the equations of motion.
In the typical situation for $SU(2)$ gauge group
only the field $\psi_{\bar{m}}$ has a zero mode.
The equations of motion for $\psi_{\bar{m}}$ read
\beq
D_{\bar{m}} \psi_{\bar{n}} -D_{\bar{m}} \psi_{\bar{n}} =0, \;\;\;
D^{\bar{m}} \psi_{\bar{m}} =0 .
\eeq
These equations are similar to those for the zero modes of the gauge field but
the field of matter $\psi$ can belong to an arbitrary representation $R$ of
the gauge group (not necessarily to an adjoint one).
The number of zero modes of the field $\psi$ depends on the representation
$R$ and is determined by the index theorem for a corresponding Dirac
operator.
For example for the gauge group $SU(2)$ and spinor representation $R$
there is generically a single zero mode \cite{abc}.

Let us consider the integration over quadratic fluctuations around
the anti-instanton field.
The result of such an integration provides us with a combination of
determinants of the Laplace-type operators.

Let $M$ be a compact K\"ahler manifold without boundaries and
let $G$ be is a compact, finite dimensional Lie group.
Let ${\cal D}^{p,q}$ denote the space of $C^{\infty}$ complex
$(p,q)$ forms on the K\"ahler manifold $M$ with values
in the direct product of
a $G$ vector bundle $L_G$ and a flat holomorphic vector bundle $L(\zeta)$
associated with a finite dimensional representation $\zeta$ of the
fundamental group $\pi_1 (M)$ of the manifold $M$ \cite{ray}.

Let us consider the anti-instanton gauge field $B_{\mu}$ obeying eqs.(4.4).
Then we can introduce the exterior derivatives
\beq
D: {\cal D}^{p,q} \to {\cal D}^{p+1,q} ,
\eeq
$$\bar{D}: {\cal D}^{p,q} \to {\cal D}^{p,q+1} ,$$
where $D$ and $\bar{D}$ are the covariant derivatives in the
presence of gauge field.
The \op s $D$ and $\bar{D}$ depend on a flat connection on $L(\zeta).$
Obviously,
\beq
D^2 =0, \;\;\; \bar{D}^2 =0 .
\eeq
However $\{ D, \bar{D} \} \neq 0$ for a non-trivial gauge field.
Therefore $D + \bar{D}$ is not nilpotent and hence does not generate
any real complex \footnote{Notice however that a modified real complex
can be defined \cite{atiyah}.}.

Let us now fix the background field gauge in the theory with the
Lagrangian (3.14) by introducing the following terms
\beq
L_{gf} = \frac{1}{2} {\rm Tr} (D^m A_m + D_m A^m)^2
+ {\rm Tr} c^+ D^nD_n c,
\eeq
where $A_m$ and $A^m$ are components of quantum part of the gluonic field,
while $c^+$ and $c$ stand for the ghost fields (here we used that $F^m_m =0$).

With this gauge fixing it is easy to check that the quadratic form for
the gluonic field is given by
\beq
L_{qu} = -2 \; {\rm Tr} A^n (D^m D_m A_n - [ D^m ,D_n ] A_m) .
\eeq
The differential operator in this expression can be represented
using the Hodge star operator and exterior derivatives $D$ and $\bar{D}$
introduced above as the Laplace type operator $\Delta_{1,0}$ acting on
${\cal D}^{1,0} ,$ where
\beq
\Delta_{p,q} = ^*\bar{D}^* D + D^*\bar{D}^* .
\eeq
The \op \ acting to the fermionic field $\chi_n$ reads as $D + ^* \bar{D}^*$
so that the relevant Laplace \op \ for the fermionic sector is $\Delta_{1,0} =
(D + ^* \bar{D}^*)^2 .$

Integrating over the non-zero modes of gluon field and fermions and
over ghosts in the
one-loop approximation we get the following ratio of
the determinants of Laplace operators
\beq
Z = \det \Delta_{0,0} \det \; '\Delta^{-1/2}_{1,0} ,
\eeq
where $\det '$ stands for the determinant without zero eigenvalues.
Here $\det \Delta_{0,0}$ corresponds to integration over ghost fields,
and $\det \; '\Delta^{-1/2}_{1,0}$ comes from integration over
fermions ($\det \; '\Delta^{1/2}_{1,0}$) and gluons
($\det \; '\Delta^{-1}_{1,0}$).
It is worth emphasizing that the boundary conditions for fermionic fields
are changed as compared to those in the untwisted SUSY theory and
are the same as in the bosonic sector.

In the presence of non-flat external metric the non-zero modes
of the fields of bosonic and fermionic fields of SUSY multiplet are
not paired up.
Therefore the dependence on metric of the ratio of determinants in eq.(4.15)
can be non-trivial.
This dependence can come also from integration over regulator fields.
The corresponding contribution is an anomaly which can spoil the BRST
invariance at the quantum level.

In the case of vanishing gauge field this combination (4.15) is given by
the Ray-Singer torsion $T_2 (M, \zeta)$
\cite{ray} ($Z =T_2^{{\rm dim}G}$)
for the Dolbeault cohomology complex defined as
\beq
\log T_p (M ,\zeta) = \frac{1}{2} \sum^2_{q=0} (-1)^q q
T_{p,q} (M,\zeta) ,
\eeq
where
\beq
T_{p,q} (M,\zeta) = \int^{\infty}_{\epsilon}
\frac{dt}{t} \; {\rm Tr} \left( e^{t\tilde{\Delta}_{p,q}} -
P_{p.q} \right) ,
\eeq
where $\tilde{\Delta}_{p,q} = ^* D^* \bar{D} + \bar{D} ^* D^*$ and
$\epsilon \to 0$ stands for a parameter of the ultraviolet cutoff.
One of the important properties of the Ray-Singer torsion is that the ratio
$T_p (\zeta_1) /T_p (\zeta_2)$ does not change under purely K\"ahlerian
deformations of the external K\"ahler metric,
i.e. under deformations which do not change its K\"ahler class \cite{ray}.

We want now to analyze the dependence of $Z$ on the external
K\"ahler metric
in the presence of non-trivial (anti-instanton) gauge field
\footnote{The idea of a possibilty of a generalization of the formalizm
of Ray and Singer in the presence of instanton field was suggested by
A.Lossev}.
To this end let us consider a more general gauge field $B_{\mu}$
obeying $F_{mn}=0$ and $F_{\bar{m} \bar{n}}=0$
while the value $g^{m\bar{m}} F_{m\bar{m}}$ can be non-zero.
Then we can still define two cohomology complexes with the
exterior derivatives $D$ and $\bar{D}$, respectively,
since these derivatives are nilpotent.
It is crucial that in this case the external gauge field
can be assumed to be independent of external metric.

We define a generalized torsion $\log \hat{T}_p (M,\zeta ,B, Ad)$
in the presence of the gauge field by the same formula as for the case
of vanishing gauge field ($Ad$ stands for the
adjoint representation of the gauge group).
For our problem we are interested in $\hat{T}_2 .$
Indeed $\hat{T}_2 = \det \tilde{\Delta}_{2,2} /
(\det \; ' \tilde{\Delta}_{2,1})^{1/2}$.
By duality we have $^* \tilde{\Delta}_{2,2}^* =\Delta_{0,0}$ and
$^* \tilde{\Delta}_{2,1}^* =\Delta_{1,0} .$
Hence $\hat{T}_2 = Z .$

Let us consider a variation of this torsion under a smooth deformations of
the metric $g_{\mu\nu} \to g_{\mu\nu} + \delta g_{\mu\nu} .$
Using a nilpotence of the operators $D$ and $\bar{D}$ it is easy
to check \cite{ray} the following identity
\beq
\delta \sum^2_{q=0} (-1)^q q T_{p,q} =
\sum^2_{q=0} (-1)^q \int^{\infty}_{\epsilon}
\frac{d}{dt} {\rm Tr} (\alpha e^{t\tilde{\Delta}_{p,q}}) dt =
\eeq
$$=\sum^2_{q=0} (-1)^q
{\rm Tr}(\alpha e^{t\tilde{\Delta}_{p,q}})|^{\infty}_{\epsilon} ,$$
where
\beq
\alpha = *^{-1} \delta *
\eeq
and $*$ is Hodge star.
{}From the above equation we get
\beq
\delta \hat{T}_p = - \frac{1}{2}\sum^2_{q=0} (-1)^q
{\rm Tr}(\alpha e^{\epsilon \tilde{\Delta}_{p,q}}) +
\frac{1}{2} \sum^2_{q=0} (-1)^q {\rm Tr} (\alpha P_{p,q}) .
\eeq
Here
$P_{p,q}$ is a projector to $(p,q)$ zero modes.
For $(1,0)$ zero modes (which are related by duality to zero modes of
$\tilde{\Delta}_{2,1}$) we have
\beq
P_{m\bar{m}}(x,y)
= A^i_m (x) P^{-1}_{ij} \bar{A}^j_{\bar{m}} (y) \sqrt{g(y)}
\eeq
where
\beq
P^{ij}_{p,q} =
\int d^4 x \sqrt{g} g^{m\bar{m}} {\rm Tr} A^i_m \bar{A}^j_{\bar{m}}
\eeq
and $A^i_m$ and $\bar{A}^j_{\bar{m}}$ stand for zero modes of
the Laplace operator in external gauge field.
Notice that $P_{m\bar{m}}$ is a matrix in the adjoint representation of the
gauge group.
As it was mentioned above we assume that there are only zero modes for
(1,0) and (0,1) sectors (and by duality in the (2,1) and (1,2) sectors)
while there are no zero modes in sectors (0,0), (0,2) and (2,0).

The first term in eq.(4.20) is an ultraviolet contribution
while the second one corresponds to the infrared domain.
This formula is analogous to the anomalous equation for the divergence
of the axial current where the ultraviolet contribution
corresponds to the axial anomaly while the infrared
contribution comes from the `soft' divergence induced by the
mass terms for fermions.

The anomalous ultraviolet contribution in eq.(4.20) can be represented
as follows
$$\delta S_{gr} + V_{mix} [A_m ,A_{\bar{m}}, g_{m\bar{m}}] .$$
Here $\delta S_{gr} (g)$ does not depend on the gauge field.
It is easy to see that $\zeta$ does not enter this ultra-violet
contribution because the corresponding connection on $L(\zeta)$ is flat.
(A non-trivial dependence on $\zeta$
can appear only from infrared contributions).
It is obvious that $S_{gr}$ is proportional to the dimension of representation
of fields over which we integrate in the path integral.
In the present case this is adjoint representation.
We denote $S_{gr}$ as ${\rm dim} G \; \log T_p (0,M) .$
If there is no zero modes in the absence of the gauge field and for
a trivial representation of the fundamental group
(the corresponding connection equals to zero) then $T_p (0,M)$ is
a particular Ray-Singer torsion.

If we normalize the path integral dividing it by the same path integral
without external gauge field then we get the following factor which does not
depend on the gauge field
$$\left( \frac{T_2 (0,M)}{T_2 (\zeta ,M)} \right)^{{\rm dim G}} .$$
It can be shown that
this factor depends only on the K\"ahler class of the metric \cite{ray}.

The second part of the variation $V_{mix} [A_m ,A_{\bar{m}}, g_{m\bar{m}}]$
is a local functional of the external gauge field
because this contribution is determined by ultra-violet contributions.
Therefore its form can be determined by a
calculation for a gauge field for which the Laplace operator
has no zero modes, in particular for small values of the gauge field
(however we still have to assume that $(2,0)$ and
$(0,2)$ components of the strength vanish).
It is useful to compare this situation with
the supersymmetric version of the theory (with twisted covariant derivative)
where two anomalies are present.
The first one is a conformal anomaly which is proportional
to $\int {\rm Tr} ^* F \wedge F \log g$ while the second anomaly appears
due to coupling of the anomalous axial current to the external vector
field $V_{\mu} .$
The axial anomaly gives the following term in the effective action
$\log Z$
\beq
\frac{1}{16\pi^2} \int {\rm Tr}_{Ad} F \wedge F \frac{1}{\Delta}
D_{\mu} V_{\mu}= \frac{1}{32\pi^2}
\int {\rm Tr}_{Ad} F\wedge F \log e/\bar{e} .
\eeq
where we used that locally the vector field $V$ is a total derivative.
Here the trace ${\rm Tr}_{Ad}$ is taken in an adjoint representation while
the generators $t^a$ of the gauge group are normalized as
${\rm Tr} t^a t^b = C_G \delta^{ab} .$

For the twisted theory formulated in terms of sections of
holomorphic and anti-holomorphic vector bundles over $M$ the
anomaly can be directly calculated by integration of
the ultraviolet contribution in eq.(4.20).
For $p=2$ we get the following expression for $V_{mixed}$
(notice that $\alpha$ vanishes in the sector of (2,0) forms)
$$ -\frac{1}{16\pi^2} {\rm Tr}_{Ad}
[g^{n\bar{m}}\delta g_{k\bar{m}}(-F^{k\bar{l}}
F_{\bar{l} n} + F_p^p F^k_n) -
g^{m\bar{m}} \delta g_{m\bar{m}} (-F^{k\bar{l}}
F_{\bar{l} k} + (F_p^p)^2)] .$$
In general it is not clear whether this anomaly integrable for an arbitrary
variation of metric.
Let the variation of metric be purely K\"ahlerian, i.e.
$\delta g_{m\bar{m}} = \partial_m \omega_{\bar{m}} + \partial_{\bar{m}}
\omega_m$ where $\omega_m$ and $\omega_{\bar{m}}$ are (1,0) and (0,1)
forms so that the K\"ahler forms $J$ and $J+ d\omega$ belong to the same
cohomology class in $H^2 (M,R) .$
Then this variation can be integrated and we get the following expression
for an anomalous contribution to the effective action $\log Z$
\beq
-\frac{1}{32\pi^2} \int  {\rm Tr}_{Ad} F \wedge F \log g.
\eeq
The anomaly in this form takes into account the change of the path integral
measure when we translate the supersymmetric theory into the twisted one.

For external anti-instanton field this term in the effective action obviously
mixes the dependence of metric and of moduli of instanton.
Hence to get a topological theory we have to cancel this anomaly by
contributions of matter in an appropriate representation of the gauge group.

Let us consider now the infrared contribution.
It is given by an expression
$$ {\rm Tr}_{Ad}
\int \sqrt{g} (g^{n\bar{m}}\delta g_{k\bar{m}} - \delta^n_k g^{m\bar{m}}
\delta g_{m\bar{m}}) P^k_n (x,x) .$$
This part of variation is also integrable.
To show that let us demonstrate that the zero modes
obey the following conditions
\beq
DA^i =0 ,\;\;\; \bar{D} \bar{A}^i =0 ,\;\;\;
^* \bar{D}^* A^i =0,\;\;\; ^* D^*\bar{A}^j =0 .
\eeq
where the indices $i,\; j$ label zero modes of the gauge field.

The equation for zero mode can be read off from the quadratic form
for fluctuations near the background field $B_{\mu} .$
We have
\beq
(^* \bar{D}^* D + D^* \bar{D}^* )A^i =0.
\eeq
Let us split the wave function for zero mode as follows
\beq
A^i = \tilde{A}^i + D \phi^i ,
\eeq
where $\phi^i$ is a scalar field and
\beq
^* \bar{D}^* \tilde{A}^i =0 .
\eeq
Then we easily get from eq.(4.26)
\beq
^* \bar{D}^* D \phi^i =0.
\eeq
Since we assume that there are no non-trivial scalar zero modes
(we consider the gauge fields which are relatively close
to the anti-instanton field with respect
to a natural metric on the space of gauge fields
while the anti-instanton field
is assumed to be irreducible) we immediately get $\phi =0 .$
In turn the field $\tilde{A}^i_n$ obeys the following equation
\beq
^* \bar{D}^* (D \tilde{A}^i) =0 .
\eeq
Acting to the left hand side of this equation by
the \op \ $D$ we get
\beq
\Delta_{2,0} (D\tilde{A}^i) =0 ,
\eeq
which is equivalent to the \cn \ that $\tilde{A}^i$ is closed
\beq
D \tilde {A}^i =0
\eeq
because we assumed that there is no zero modes in the ${\cal D}^{2,0}$ space.
Thus $A^i =\tilde{A}^i .$
Similarily one can show that $\bar{D} \bar{A}^i = ^* D^* \bar{A}^i =0 .$
Thus we can say that $A^i$ and $\bar{A}^i$ are ``harmonic''.

Let us now consider the variation in metric of the \cn s $D^n A^i_n =0$
and $D_m A^i_n - D_n A^i_m =0 .$
They read
\beq
D^n \delta A^i_n + \delta g^{m\bar{m}} D_{\bar{m}} A^i_m =0 ,
\eeq
$$D_m \delta A^i_n - D_n \delta A^i_m =0 .$$
Notice that in the case of K\"ahler metric we should not vary
the covariant derivatives because the corresponding gravitational
connection vanishes and the external gauge field is assumed to be
independent of metric.
The solution to these equations with respect to $\delta A^i_n$ is given by
\beq
\delta A^i_n = D_n \Phi^i + c^i_j A^j_n ,\;\;\;
\delta A^i_{\bar{n}} = D_{\bar{n}} \bar{\Phi}^i + \bar{c}^i_j A^j_{\bar{n}} .
\eeq
Here $\Phi^i$ and $\bar{\Phi}^j$ are complex (conjugated to each other)
scalar fields,
\beq
\Phi^i = -\frac{1}{D^n D_n} \delta g^{m\bar{m}} D_{\bar{m}} A^i_m ,
\eeq
and $c^i_j$ and $\bar{c}^i_j$ are the components of a (real) connection
on the bundle ${\cal K}\times {\cal G}$
where ${\cal K}$ and ${\cal G}$ are the spaces of
zero modes of gauge field and of K\"ahler metrics
consistent with the complex structure on $M,$
respectively.
An integrability \cn \ for eqs.(4.34) is equivalent to
\beq
{\cal D} \phi =0 ,\;\;\;
[{\cal D},{\cal D}] =0,
\eeq
where ${\cal D}^i_j = \delta^i_j \delta - c^i_j .$
Hence $c^i_j$ is a flat connection
and can be locally included into normalization of zero modes.
Thus
the infrared variation of the torsion can be locally represented as follows
\beq
{\rm Tr}_{Ad} (\alpha P_{1,0}) = \delta \log \det P^{ij}_{1,0} .
\eeq
Therefore this infrared variation is locally integrable
and we get for the torsion the following expression
\beq
Z= \frac{\hat{T}_2 (M,\zeta ,B,Ad)}{\hat{T}_2 (M,\zeta ,0,Ad)} =
\eeq
$$Y_G\; \left( \frac{T_2 (0 ,M)}{T_2 (\zeta ,M)} \right)^{{\rm dim} G}
(\det P)^{-1/2} \exp (-\frac{1}{32\pi^2} \int {\rm Tr}_{Ad}
F\wedge F\log g ) .$$
Here $Y_G$ is a factor which does not depend on external metric $g.$
The factor $(T_2 (0 ,M)/T_2 (\zeta ,M))^{{\rm dim} G}$ does not change
under smooth deformations of the K\"ahler metric which do not change its
K\"ahler class.
If the infrared contribution is not globally integrable
this torsion is a section of a line bundle over ${\cal G} .$

Let us consider now the sector of fields of matter in the representation $R$
of the gauge group.
For the fields of matter in the semiclassical
approximation get the following expression
\beq
Z_R = \det \tilde{\Delta}_{0,0}^{-1} \det\; '\tilde{\Delta}^{1/2}_{0,1}=
\det \Delta_{0,0}^{-1} \det\; '\Delta^{1/2}_{1,0}.
\eeq
Here we used the fact that we can represent the determinant $\det '
\; \Delta_{1,0}$
as a path integral over (1,0) and (0,1) forms with a bilinear action (with
operator $\Delta_{1,0}$ in the quadratic form) and integrate by parts.
Then we get an \op \ $\tilde{\Delta}_{0,1}$ while
no boundary terms appear under integration by parts
on a compact manifold without boundaries.
Hence their determinants coincide.
The same argument can be used for (0,0) sector.

Thus we can represent this ratio of determinants as a generalized torsion
$\hat{T}_2 (M,\zeta ,B,R)$ for the representation $R.$
It is defined by eq.(4.16) with the covariant derivatives in
the representation $R.$
By the same arguments as for the adjoint representation
one can show that the ultraviolet and infrared
contributions to the variation in metric of $Z_R$ are integrable
separately.
Thus we get for $Z_R$ the following expression
\beq
Z_R = \frac{\hat{T}_2 (M,\zeta ,0,R)}{\hat{T}_2 (M, \zeta ,B,R)} =
f_R \; \left( \frac{T_2 (\zeta ,M)}{T_2 (0 ,M)} \right)^{{\rm dim} R}
(\det \tilde{P})^{1/2} \exp(\frac{1}{32\pi^2}
\int {\rm Tr}_R F\wedge F\log g ) ,
\eeq
where the trace ${\rm Tr}_R$ is taken in the representation $R$
while the generators $t^a$ of
the gauge group are normalized as ${\rm Tr} t^a t^b = C_R \delta^{ab} ;$
$f_R$ is a factor independent on external metric,
${\rm dim}R$ is a dimension of a representation of the
multiplet of matter (it can be reducible)
and $\tilde{P}$ is a matrix of bilinear integrals of
fermionic zero modes (we assume that there is no scalar field zero modes,
i.e. the instanton field is irreducible).
In the simplest case when only (0,1) zero modes exist we have
\beq
\tilde{P}^{IJ}
= \int d^4 x \sqrt{g} g^{m\bar{m}} <\bar{\psi}^I_m \psi^J_{\bar{m}}> ,
\eeq
where $\psi^J_{\bar{m}}$ is a zero mode wave function (labelled by
an index $J$)
while $\bar{\psi}^I_m$ is a complex conjugated zero mode.

Notice that though $Z$ and $Z_R$ are the ratios of the generalized
torsions they can depend on a representative of the K\"ahler class
of external metric in contrast to the case of absence of external gauge field
\cite{ray}.
The point is that the argument of Ray and Singer in ref.\cite{ray}
is essentially based on a nilpotence of the \op \
$D +\bar{D}$ while it is not nilpotent in the presence of external gauge field.

The total contribution of non-zero modes reads
$$ZZ_R = \frac{\hat{T}_2 (M,\zeta ,0,R)}{\hat{T}_2 (M, \zeta ,B,R)}
\frac{\hat{T}_2 (M,\zeta ,B,Ad)}{\hat{T}_2 (M, \zeta ,0,Ad)} .$$
Now we can see that the mixed anomaly is cancelled if
\beq
C_G -C_R = C_G -\sum_i C_{R_i} =0,
\eeq
where $R=\sum_i R_i ,$ $R_i$ are irreducible representations of the gauge
group.
The condition of cancellation of the Ray-Singer torsion is the following
\beq
{\rm dim} G - \sum_i {\rm dim} \; R_i  = 0.
\eeq
However since the Ray-Singer torsion does not depend on the gauge field
and can depend only on the K\"ahler class of the metric
we do not impose the \cn \ (4.43).
Thus the whole physical correlator does not change under
smooth variations of external Kahler metric which do not change its K\"ahler
class if the mixed anomaly is cancelled.
The dependence on the K\"ahler class is factorized out into a power of
a ratio of two Ray-Singer torsions.

The ultarviolet contribution to the generalized
torsions has been shown to be integrable under an assumption
that the variation of metric is purely K\"ahlerian, i.e. it does not change
the K\"ahler class of metric.
Actually it is easy to check that the cancellation of a mixed anomaly
does not depend on this assumption if the \cn \ (4.42) is satisfied.

There is also another restriction to the sector of matter:
the theory should not have (both local and global) gauge anomalies.

Notice that
if the multiplet of matter is in the adjoint representation of the gauge group
($C_G =C_R$) both the anomaly and the infrared contributions are cancelled
in the product $Z Z_R$ and we get Witten's theory.

Now we can substitute the anti-instanton field into the expressions
for $Z$ and $Z_R$ and study the physical correlators.
If $t^i$ are coordinates on the moduli space the wave functions
for zero modes of gauge field can be represented as follows \cite{harvey}
\beq
A^i_{\mu} = \partial^i B_{\mu} + D_{\mu} \epsilon^i ,
\eeq
where $\epsilon (t^i ,x)$ are gauge parameters ($x \in M$), $\partial^i =
\partial /\partial t^i$ and $B_{\mu}$ stands for anti-instanton field.
This parameter $\epsilon^i$ plays a role of a natural connection on
the moduli space ${\cal M}$ \cite{harvey}.

It is useful to introduce a natural metric on
${\cal M}$ which is induced by a metric on $M $ \cite{itoh}:
\beq
\hat{G}^{ij} = \int_M \sqrt{g} {\rm Tr} A^i_{\mu} A^{\mu j}.
\eeq
It is convenient to define also the K\"ahler form
\beq
\hat{J}^{ij} = \int d^4 x \sqrt{g} J^{\mu\nu} {\rm Tr} A^i_{\mu}
A^j_{\nu} .
\eeq
Moreover we can define a natuarl complex structure on ${\cal M}$
induced by the complex structure on $M :$
\beq
\hat{J}^i_{\; j} = \hat{J}^{il} \hat{G}_{l j} .
\eeq
It is easy to check that there is the following relation between
the complex structures on $M$ and ${\cal M}$
\beq
\hat{J}^i_j A^j_{\mu} = J^{\nu}_{\mu} A^i_{\nu} .
\eeq
This means that the only non-vanishing components of the wave functions
of zero modes for the gauge field are $A^i_m$ and
$\bar{A}^{\bar{k}}_{\bar{m}} ,$ where the indices $i$ and $\bar{k}$ are
holomorphic and anti-holomorphic, respectively.
{}From now on we shall use the letters $i,\; j,...$ for holomorphic
and $\bar{k} ,\; \bar{l} ...$ for anti-holomorphic indices on ${\cal M} .$

The metric on the moduli space turns out to be K\"ahler one so that
in the well adapted frame the metric has only components with one
holomorphic ($i$) and one anti-holomorphic indices ($\bar{k}$)
\beq
\hat{G}^{i\bar{k}}=P^{i\bar{k}} = \int d^4 x \sqrt{g} g^{m\bar{m}}
{\rm Tr} A^i_m \bar{A}^{\bar{k}}_{\bar{m}} .
\eeq
With respect to this complex structure the coordinates on the moduli space
can be splitted into holomorphic and anti-holomorphic ones,
$t^i$ and $\bar{t}_{\bar{k}}$ respectively.
The wave functions for zero modes can be represented as follows
$A^i_n = \partial^i B_n + D_n \epsilon^i$ and
$\bar{A}^{\bar{k}}_{\bar{n}} = \partial_{\bar{k}} \bar{B}_{\bar{n}} +
D_{\bar{n}} \epsilon^{\bar{k}} ,$ where $B_{\mu} =(B_n ,\bar{B}_{\bar{n}})$
is the anti-instanton field.
It follows from eq.(4.48)
that $\partial^{\bar{k}} B_n = - D_n \epsilon^{\bar{k}}$ and
$\partial^i \bar{B}_{\bar{n}} =- D_{\bar{n}} \epsilon^i .$

The K\"ahler form $\hat{J}^{i\bar{j}}$ is closed.
This can be easily seen if we locally (on ${\cal M}$) fix the gauge
of the anti-instanton field by a \cn \ $\epsilon^i =\epsilon^{\bar{k}} =0 .$
Then $\hat{G}^{i\bar{k}} = \partial^i \partial^{\bar{k}} \int_M \sqrt{g}
{\rm Tr} B_n \bar{B}^n .$
However in general the K\"ahler potential $\hat{K}$ ($\hat{G}^{i\bar{k}}=
\partial^i \partial^{\bar{k}} \hat{K}$)
is a non-trivial functional of the instanton field \cite{harvey}
\beq
\hat{K} = \frac{1}{2} \int_M {\cal F} \wedge J - \log \int_M J\wedge J ,
\eeq
where ${\cal F}$ is a solution to the following equation
\beq
{\rm Tr} F\wedge F = i \partial \bar{\partial} {\cal F} .
\eeq
Here $\partial$ and $\bar{\partial}$ are the Dolbeault \op s on $M.$

It is easy to see that the K\"ahler class of $\hat{J}^{i \bar{k}}$
depends only on the K\"ahler class of the K\"ahler form $J$ on $M .$
Indeed in the case of
K\"ahlerian variation of metric $\delta J = \partial \bar{\omega}
+ \bar{\partial} \omega =d\Omega $
($\omega$ and $\bar{\omega}$ are (1,0) and (0,1)
forms on $M$, respectively) we have
\beq
\delta \hat{J}^{i\bar{k}} = \partial^i \hat{\bar{\omega}}^{\bar{k}} -
\partial^{\bar{k}} \hat{\omega} ,
\eeq
where
\beq
\hat{\bar{\omega}}^{\bar{k}} =
\int_M (J\wedge {\rm Tr} \delta B \wedge \partial^{\bar{k}} B
+ \Omega \wedge {\rm Tr} F\wedge \partial^{\bar{k}} B) ,
\eeq
$$\hat{\omega}^i = \int_M ( J\wedge {\rm Tr} \delta B \wedge \partial^i B +
\Omega \wedge {\rm Tr} F\wedge \partial^i B) .$$
In eqs.(4.53) we used the total field $B_{\mu} =(B_n ,\bar{B}_{\bar{n}}) .$
Here $\delta B_{\mu}$ is a variation of $B_{\mu}$ in metric and
we used that the K\"ahler form $\hat{J}$ (eq.(4.46)) can be
rewritten as follows
$$\hat{J}^{i\bar{k}} = \frac{1}{2} \partial^i \int \sqrt{g} J^{\mu\nu}
{\rm Tr} B_{\mu} \partial^{\bar{k}} B_{\nu} -
\frac{1}{2} \partial^{\bar{k}} \int \sqrt{g} J^{\mu\nu}
{\rm Tr} B_{\mu} \partial^i B_{\nu} ,$$
and the condition $J\wedge F=0 .$

It is worth noticing that the connection $\epsilon^i$ ($\epsilon^{{\bar k}}$)
$\;$is holomorphically (anti-holomorphically) flat.
Indeed with the complex structure on ${\cal M}$ defined above we can see
that the integrability condition for the equation
$\partial^i \bar{B}_{\bar{n}} = - D_{\bar{n}} \epsilon^i$ reads
\beq
\bar{D} (\partial^i \epsilon^j -\partial^j \epsilon^i +i [ \epsilon^i ,
\epsilon^j]) =0.
\eeq
This equation is equivalent to
\beq
\partial^i \epsilon^j -\partial^j \epsilon^i +i [ \epsilon^i ,
\epsilon^j] =0
\eeq
because we assume that the anti-instanton field is irreducible.
Moreover
$$ (\partial^i +i \epsilon^i) \bar{A}^{\bar{k}} = \bar{D} \Phi^{i\bar{k}}
,\;\;\;
(\partial^{\bar{k}} + i\epsilon^{\bar{k}}) A^i = -D \Phi^{i\bar{k}} ,$$
where $\Phi^{i\bar{k}} = \partial^i \epsilon^{\bar{k}} -
\partial^{\bar{k}} \epsilon^i +i[\epsilon^i ,\epsilon^{\bar{k}}] .$
Notice that the strength $\Phi^{i\bar{k}}$ can be non-zero.

We can now write down the integration measure on the moduli space ${\cal M}$
for the correlator.
Taking the measure $\prod_{x,\mu ,a} d A^a_{\mu} (x)$
on the space of gauge connections modulo gauge transformations
we get an induced measure on the moduli space
\beq
\prod_i d t_i \prod_{\bar{k}} d t_{\bar{k}} \det P ,
\eeq
where the matrix $P^{i\bar{k}}$ is evaluated for the
anti-instantonic background field
and corresponds simply to the Jacobian of the change of variables
in the path integral.
However there is a contribution from integration over non-zero modes.
We actually calculated it through its variation in metric.
If the \cn \ of cancellation of anomaly fulfilled then
the correct measure reads
\beq
\prod_i d t_i \prod_{\bar{k}} d t_{\bar{k}}
(\det \tilde{P} \det P)^{1/2} \left(
\frac{T_2 (0 ,M)}{T_2 (\zeta , M)} \right)^{{\rm dim} G - {\rm dim} R}
Y_G f_R
\eeq
We take into account here that the classical action equals to zero for
instanton configuration.
The integrand eq.(4.57) stands for a generalization of
square root of a determiant of real metric on the moduli space.

The physical correlator is given now by an integral over ${\cal M}$ of
a product of the operators
where the quantum fields are replaced by the normalized zero modes
(a normalization factor comes from the Jacobian of a change of variables
for fermionic zero modes in the fermionic path integral)
and the instanton measure.
Let us consider for example the gauge group $SU(2)$ and
the matter in the representation given by four copies of
a spinor representation.
Then the mixed anomaly is cancelled because $C_G =2$ and $C_R = 4\times 1/2
=2,$
while the whole correlator is
proportional to $T_2 (\zeta , M)/T_2 (0 ,M) .$
The fermions of matter have altogther 4 zero modes which are (at least naively)
(0,0) forms on the moduli space
(generically \cite{abc} each fermionic doublet $\psi_{\bar{n}}$
has a single zero mode).
We take a physical operator for the sector of matter defined by eq.(3.25)
with a non-degenerate antisymmetric matrix $a_{IJ}$
and for the sector of the gauge multiplet we take
$O^{(0)}_{mn} (x_1)  O^{(0)}_{kl} (x_2)$ ($x_{1,2}$ are coordinates on $M$).
Then the physical correlator reads as follows (we substitute the zero modes
$A^i_n$ for the fields $\chi_n$)
\beq
\frac{T_2 (\zeta ,M)}{T_2 (0, M)}
\int_{{\cal M}} {\rm Tr} (A_m \wedge A_n) (z_1) \wedge
{\rm Tr} (A_k \wedge A_l) (z_2)
(\int E \wedge \psi^0 \wedge \psi^0)^2 Y_G f_R  ,
\eeq
where $\psi^0$ is a zero mode wave function of fermionic field $\psi^I$
(the same for all 4 copies of a spinor representation of the gauge group).
Notice that the wave functions for zero modes in eq.(4.58) are not normalized
because they absorbed the factor $(\det P \det \tilde{P})^{1/2} .$

With the complex structure on ${\cal M}$ the \op \ constructed of
the fields $\chi_n$ becomes a (4,0) form on the moduli space
(notice that $A^i$ are (1,0) forms on ${\cal M}$).
To keep the covariance on the moduli space with respect to the K\"ahler metric
we need to get for a correlator a volume (4,4) form integrated over all
the moduli space.
That means that the product of zero modes of the fields of matter
times $Y_G f_R$
should transform as (0,4) form on the moduli space.
However it is not clear in general case of an arbitrary irreducible
representation of the gauge group how do the zero modes of matter transform
with respect to the moduli space, i.e. whether they are forms or
section of a spinor bundle over ${\cal M} .$
The number of zero modes in general case is given by the index theorem
while an unravelling of the geometrical properties of these zero modes
is an interesting problem.

\section{Heterotic topological model as a deformation of Witten's theory}
\setcounter{equation}{0}

It is possible however to show that the integrand in eq.(4.58)
is indeed a (4,4) form on the moduli space.
To this end we shall deduce the same result using a different
ultraviolet regularization.
This approach will be also more transparent for the question of
dependence of the physical correlator on external metric.
Let us start with Witten's theory which of course does not have any anomalies.
It is equivalent to our theory with matter in the adjoint representation
of the gauge group.
We can consider a deformation of such a theory by the mass term \op \
\beq
\Lambda \{ Q, \int S\wedge {\rm Tr} \bar{\psi} \bar{\phi} \} +
\Lambda \int E\wedge {\rm Tr} (\psi \wedge \psi + N \phi) ,
\eeq
where $S$ is an arbitrary (non-singular) (0,2) form, $E$ is a holomorphic
(2,0) form and $\Lambda$ is a parameter (a mass).
The fields of matter are in an adjoint representation of the gauge group.
This deformation is $Q$-closed and does not spoil $Q$ invariance of the theory.

In Witten's theory the physical correlator with two insertions of
this operator can be interpreted up to a constant factor as a
correlator without such insertions in a deformed theory with the mass
\op \ for the fields of matter (eq.(5.1))
with small coefficient (we change $\Lambda \to m$ in eq.(5.1)) $m\to 0 .$
In semiclassical approximation the deformed theory with $\Lambda \to \infty$
differs from Witten's one by a factor which is a ratio of two path integrals
for the fields of matter with masses $\Lambda$ and $m,$
$Z(\Lambda)/Z(m) ,$ where $Z(m)$ and
$Z(\Lambda)$ stand for these two path integrals.
It is clear that such a ratio is well defined since the fields with
mass $\Lambda$ can be understood as the regulator fields for the fields with
mass $m.$
In turn one can see that when evaluated in the external anti-instanton field
this ratio is (0,0) form on the moduli space because we integrate over
zero modes both in the nominator and in the denominator.
Furthermore the ratio $Z(\Lambda)/Z(m)$
does not depend on external metric because the
variation in metric of both non-deformed and deformed lagrangians is $Q$-exact.
However this ratio depends on a choice of (2,0) form $E .$

The variation of $\log Z(\Lambda)/Z(m)$ in the
(2,0) form $E$ can be easily calculated.
It is worth noticing that the form $E$ can have zeros.
Therefore for this calculation it is useful to regularize $E \to E_r$ where
$E_r$ does not have zeros, and to put $S=E^{-1} .$
At the end of calculations we have to switch off the regularization.
In the case $F_{mn} = F_{\bar{m}\bar{n}} = 0$ we get for the variation
$$\frac{1}{32\pi^2} {\rm Tr}_{Ad} (F^{m\bar{m}} F_{\bar{m} m} -
(F^n_n)^2) \delta \log \det E - \frac{1}{2} \delta \log \det \hat{E} .$$
Here
\beq
\hat{E}^{\bar{k}\bar{l}} = \int E\wedge {\rm Tr}_{Ad} \psi^{\bar{k}}
\wedge \psi^{\bar{l}} ,
\eeq
and $\psi^{\bar{k}}$ stands for a normalized wave function for $\bar{k}$
zero mode of fermionic field of matter (the \zm s are normalized by
a factor $(\det \int {\rm Tr} ^* \bar{\psi}^{\bar{k}} \psi^{\bar{l}})^{1/2}$
where $\bar{\psi}^{\bar{k}}$ is complex conjugated to $\psi^{\bar{k}}$).
Notice that in the case of the anti-instanton background
$\psi^{\bar{k}}$ is antiholomorphic vector on ${\cal M} .$
The first term in eq.(5.2) is the ultraviolet contribution from the regulator
fields (with mass $\Lambda$) while the second one
corresponds to the contribution of fields with mass $m.$
Notice that in the limit $m \to 0$ only zero modes contribute to
the infrared part of the variation.

This variation can be integrated
up to a factor $\tilde{f}_G$ which does not depend on $E$.
Thus we get
$$Z(\Lambda)/Z(m) =$$
\beq
= \tilde{f}_G (\tilde{T}_2 (0 ,E,M))^{{\rm dim} G}
\; (\det \hat{E})^{-1/2}
\exp \left( -\frac{1}{32\pi^2} \int {\rm Tr}_{Ad}
F\wedge F \log \det E \right) .
\eeq
The factor $\tilde{T}_2 (0 ,E,M)$ is a ``torsion'' which does not depend
on external gauge field but can be a non-trivial functional of $E .$
If we normalize the path integral to the same path integral without gauge field
we get the same factor $\;$$(T_2$$ (0,M)/T_2 (\zeta ,$$M))^{{\rm dim} G}$
as in eq.(4.38) because the dependence on $E$ of
the path integral in absence of the gauge field is given by the same
ultraviolet contribution as $\tilde{T}_2 $$(0,$$E,$$M) .$

One can see that the expression (5.3) is explicitly invariant under
rescaling of the (2,0) form $E.$
This is a manifestation of the fact that this ratio $Z(\Lambda)/Z(m)$
is (0,0) form.

Let us now analyse a dependence of the ratio (5.3) on a choice of
the form $E .$
For the case of hyperk\"ahler there is a canonical (2,0) form which is
not degenerate everywhere and can be used to construct the mass terms.
However for an arbitrary K\"ahler manifold the form $E$ can have zeros.
Therefore we shall consider a regularized form $E$ which
is non-zero everywhere but it can be non-holomorphic.
Then the theory is not $Q$-invariant due to mass terms.
Let us consider an arbitrary non-singular (not necessarily
holomorphic) variation $\delta E$ of $E.$
The corresponding variation of the mass term can be removed by an
appropriate simultaneous change of the variables of integration in the
path integrals (both for physical and regulator fields):
the fields $\psi_{\bar{m}},$ $\phi$ and $N_{\bar{m}\bar{n}}$
are rescaled by $1+ h ,$ where $h =- (E^{-1})^{mn}\delta E_{mn} ,$
while other fields are rescaled by $1-h$ (of course we have to consider
the dependence of $h$ in linear approximation).
Then instead of the lagrangian with a variation of the mass term we get
the lagrangian with an additional deformation
\beq
\{ Q,\int \sqrt{g} \partial_m h {\rm Tr}\bar{\phi} \psi^m  \} +
\int \sqrt{g} \partial_{\bar{m}}h {\rm Tr} (\bar{\psi}^{\bar{m}\bar{n}}
\psi_{\bar{n}} - \phi D^{\bar{m}} \bar{\phi}) .
\eeq
The first term in the above expression is $Q$-exact while the second one
is not even $Q$-invariant.
Now let us continuously switch off the regularization
so that the forms $E$ and $\delta E$ may have zeros.
The first term above gives vanishing contributions in the physical
correlators while the second one vanishes only if $h$ is holomorphic.
That means that the ratio $Z(\Lambda)/Z(m)$ does not depend on deformations of
$E$ which do not change the positions of zeros of $E .$
For the case of hyperk\"ahler manifold with a canonical holomorphic form
$E$ without zeros or poles $Z(\Lambda)/Z(m)$ does not depend on a choice of
$E .$

Let now the deformation $h$ be meromorphic with poles.
Generically these poles determine 2D surfaces (the equation $h^{-1} = 0$
is equivalent to two real equations for 4 real variables).
Then taking into account an identity
$\bar{\partial} (1/z) = \delta^{(2)}(z),$ where $z$ is a complex variable,
we get for a variation of $Z(\Lambda)/Z(m)$
a sum of correlators with insertions of the current
$<\bar{\psi}^{\bar{m}\bar{n}}
\psi_{\bar{n}} - \phi D^{\bar{m}} \bar{\phi}>$ integrated over
2-dimensional hypersurfaces on $M .$

This analysis can be easily generalized for an arbitrary representation $R$ for
the multiplet of matter.
Again we can introduce mass terms into the lagrangian of matter
with the same (2,0) form $E$
\beq
\sum_{IJ} a_{IJ}
(\Lambda \{ Q, \int S\wedge < \bar{\psi}^I \bar{\phi}^J > \} +
\Lambda \int E\wedge <\psi^I \wedge \psi^J + N^I \phi^J >) ,
\eeq
where $a^{IJ}$ is an appropriate nondegenerate constant matrix (the indices
$I,\; J$ correspond to different irreducible multiplets of matter).
A choice of this matrix depends on the represenation $R .$
For the case when the gauge group is $SU(2)$ and there is
an even number of copies of a spinor
represenation for the fields of matter
this matrix should be antisymmetric.

One can show that the ratio of
two path integrals $Z_R (m)/Z_R (\Lambda)$
with different masses $m$ and $\Lambda$ (they are regularizing
each other) depends only on the equivalence class of $E$
(two forms $E$ are equivalent if they differ by a nonsingular holomorphic
factor).
This ratio is again a (0,0) form on the moduli space when considered
in the external anti-instanton field.
The dependence on an arbitrary variation of $E$ can be calculated
as well.
We have
$$Z_R (\Lambda)/Z_R (m) =$$
\beq
= \tilde{f}_R (\tilde{T}_2 (0 ,E,M))^{{\rm dim} R}
(\det \hat{E}_R)^{-1/2}
\exp \left( -\frac{1}{32\pi^2} \int {\rm Tr}_R
F\wedge F \log \det E \right) ,
\eeq
where $\hat{E}_R$ is a matrix of bilinear combinations of the zero mode
wave functions of fermions of matter integrated with $E.$
The factor $\tilde{f}_R$ does not depend on $E$ while
$\tilde{T}_2$$ (0 ,E,M)$ is a ``torsion'' that does not depend on
the external gauge field a representation $\zeta$
of the fundamental group of $M$ but it can be a functional of $E.$
If we normalize the path integral dividing it by the same integral without
gauge field we get $(\tilde{T}_2 (0,E,M)/\tilde{T}_2 (\zeta ,E,M))^{{\rm
dim}R}$
$= (T_2 (0,M)/T_2 (\zeta ,M))^{{\rm dim}R}$ which does not depend on $E.$

For a particular case of the gauge group $SU(2)$ with $2N$ copies of
spinor representation for the fields of matter we have
\beq
\det \hat{E}_R = (\int E\wedge \psi^{(0)} \wedge \psi^{(0)})^N .
\eeq
Notice that here the wave function $\psi^{(0)}$ is normalized ($\int \;
^* \bar{\psi}^{(0)} \wedge \psi^{(0)} =1 ,$ where $\bar{\psi}^{(0)}$ stands for
complex conjugated zero mode).

The exponential of $\int {\rm Tr}_R F\wedge F \log \det E$ in the above
expression is a manifestation of the anomaly that was already discussed
in this paper.
Here we see once again that if $C_G = C_R$ (i.e. ${\rm Tr}_{Ad} F\wedge F =
{\rm Tr}_R F\wedge F$) then
the anomaly is cancelled in the product
\beq
\left( \frac{Z(\Lambda)}{Z(m)} \right)
\left( \frac{Z_R (m)}{Z_R (\Lambda)} \right) =
\left( \frac{\det\hat{E}_R}{\det \hat{E}} \right)^{1/2}\;
\left( \frac{T_2 (0 ,M)}{T_2 (\zeta ,M)} \right)^{{\rm dim} G -{\rm dim} R}
\frac{\tilde{f}_G}{\tilde{f}_R} .
\eeq
If we now consider in Witten's non-deformed theory
a physical correlator with two insertions of the mass term operator
for the matter multiplet (in adjoint representation of the gauge group)
we can translate it into the theory where
the matter in adjoint representation is replaced by the matter in
an appropriate representation $R$ of the gauge group.
Indeed the theory has non-anomalous BRST symmetry and therefore allows
for a localization of the path integral to the anti-instanton configurations.
Therefore for such a translation
we should multiply the integrand in the integral over ${\cal M}$
for the correlator in Witten's theory by an above factor
where we substitute the anti-instanton field.
Then we get the expression for
the correlator in this heterotic theory given by the same formula
of eq.(4.58) (with some change of notations).

Notice that the ratio $Z_R (m)/Z_R (\Lambda)$
in the external anti-instanton field does not depend on the holomorphic modulii
(for any representation $R$ of the gauge group) because the BRST charge $Q$
does not depend on $A_n$ gauge field (the parameter $\epsilon^p$ which appears
for the derivative in $t_p$ enters as a gauge transformation and does not
affect the result of calculations).
Therefore we get an identity
\beq
\partial^p \log \det \hat{E}_R = \frac{1}{16\pi^2}
\partial^p \int {\rm Tr}_R F\wedge F \log \det E  + 2 \partial^p
\log \tilde{f}_R = 2 \partial^p \log \tilde{f}_R .
\eeq
It is interesting that this derivative does not depend on the
holomorphic (2,0) form $E.$

Let us consider for simplicity the case of $SU(2)$ gauge group and
spinor representation for the matter fields.
Let us consider $\partial^p \psi .$
It obeys the following equations
\beq
\bar{D} (\partial^p +i\epsilon^p) \psi =0,\;\;\;
^*D^* (\partial^p +i\epsilon^p) \psi - i ^*A^p \; ^* \psi =0.
\eeq
A general solution to this equation reads (we use a normalization of zero modes
in the adjoint representation determined by eq.(4.44))
\beq
(\partial^p + i\epsilon^p) \psi = \bar{D} u^p + c^p \psi ,
\eeq
where $u^p$ is (0,1) form on $M$ while $c^p$ are some coefficients
depending on modulii and external metric.
Substituting this expression into $\partial^p \hat{E}_R$ we get
\beq
\partial^p \hat{E}_R = 2c^p \hat{E}_R .
\eeq
Hence $\partial^p \tilde{f}_R = c^p$ and $\tilde{f}_R$
is a section of a line bundle over ${\cal M} \times {\cal G}$
while $c^p$ is flat connection on a bundle over ${\cal M} \times {\cal G}$
with a fibre given by the space of zero modes for a representation $R .$

We can now (locally) redefine the wave functions for zero modes and include
$c^p$ into their normalization.
Then $\hat{E}_R$ is locally anti-holomorphic
\beq
\partial^p \hat{E}_R =0 .
\eeq
For the zero modes of adjoint representation using the fact that
the connection $\epsilon^p$ is flat
(with a normalization of zero modes determined by eq.(4.44)) and
$(\partial^p + i\epsilon^p)\bar{A}^{\bar{k}} = \bar{D} \Phi^{p\bar{k}}$ we get
\beq
\partial^p \det \hat{E} = \partial^p \hat{f}_G =0.
\eeq
Let us now consider the variation of $Z_R (m)/Z_R (\Lambda)$ in metric.
Since the dependence of it on metric is due to a dependence of metric of
the anti-instanton field we have to consider first the variation of the
anti-instanton gauge field $B_{\mu}$ under the variation of metric.
The variation of equation $F_{mn} = 0$ gives us
\beq
D_m \delta B_n -D_n \delta B_m =0,
\eeq
while from equation $g^{m\bar{m}} F_{m\bar{m}} =0$ we get
\beq
\delta g^{m\bar{m}} F_{m\bar{m}} + g^{m\bar{m}}(D_m \delta B_{\bar{m}}
- D_{\bar{m}} \delta B_m ) =0.
\eeq
The solution to this equations reads
\beq
\delta B_n = D_n \Phi + c_i A^i_n ,
\;\;\; \delta B_{\bar{n}} = D_{\bar{n}}
\bar{\Phi} + \bar{c}_{\bar{i}} A^{\bar{i}}_{\bar{n}} ,
\eeq
where $\Phi$ and $\bar{\Phi}$ are complex conjugated to each other
scalar fields obeying the following \cn \
\beq
\Phi - \bar{\Phi} = - \frac{1}{\Delta} \delta g^{m\bar{m}} F_{m\bar{m}}
\eeq
while $\Delta = D_n D^n .$
The constant coefficients $c_i$ and $\bar{c}_{\bar{i}}$ are complex conjugated
to each other and can depend on the moduli of instanton and on metric.

The variation of the zero mode wave function
$\delta \psi^{(0)}$ under a variation of metric obeys the following equations
\beq
\bar{D} \delta \psi^{(0)} -i \delta \bar{B} \wedge \psi^{(0)} =0,
\eeq
$$^* D^* \delta \psi^{(0)} - i^* \delta B ^* \psi^{(0)}
+ (\delta ^*) D^* \psi^{(0)} + ^* D (\delta ^*) \psi^{(0)} =0.$$
A general solution to these equations reads
\beq
\delta \psi^{(0)} = i (\bar{\Phi}+ c_{\bar{p}} \epsilon^{\bar{p}}) \psi^{(0)}
+ c_{\bar{p}} \partial^{\bar{p}}
\psi^{(0)} + \bar{D} \epsilon + c \psi^{(0)} ,
\eeq
where $c$ is a section of a line bundle over ${\cal M}\times {\cal G} .$
Here we used the expression for variation in metric of the gauge field.
Then for the variation of $K$ we get
\beq
\delta \hat{E}_R = 2c \hat{E}_R + c_{\bar{p}} \partial^{\bar{p}} \hat{E}_R .
\eeq
For the adjoint representation of the gauge group we can deduce
in a similar way
\beq
\delta A^{\bar{k}}_{\bar{m}} = i({\bar{\Phi}}
+ c_{\bar{p}} \epsilon^{\bar{p}} ) A^{\bar{k}}_{\bar{m}} +
\bar{D}_{\bar{m}} \tilde{\epsilon}^{\bar{k}} + c_{\bar{p}} \partial^{\bar{p}}
A^{\bar{k}} + c^{\bar{k}}_{\bar{l}} A^{\bar{l}} ,
\eeq
where $\tilde{\epsilon}^{\bar{k}}$ is a (0,0) form on $M$ and
$c^{\bar{k}}_{\bar{l}}$ are some coefficients which can depend on the
coordinates on the moduli space and external metric.
Then we get for the variation of $\det \hat{E}$ in external metric
\beq
\delta \det \hat{E} = 2 c^{\bar{k}}_{\bar{k}} \hat{E} + c_{\bar{p}}
\partial^{\bar{p}} \det \hat{E} .
\eeq
The integrability \cn \ for eq.(5.23)
gives a linear (variational) equation which relates
$c,$ $c_{\bar{p}}$ and $\det \hat{E}_R$
\beq
(2\delta c + \delta c_{\bar{p}} \partial^{\bar{p}} c)\hat{E}_R
+ (2\delta c_{\bar{p}}\partial^{\bar{p}} c + c_{\bar{q}}
\partial^{\bar{q}} c_{\bar{p}} \partial^{\bar{p}}) \det \hat{E}_R =0,
\eeq
and a similar equation for the adjoint representation.
We see here that the vector field $c_{\bar{p}} \partial^{\bar{p}}$ on
${\cal M}$ plays a role of connection for the variation in metric.
Now we notice that the whole correlator in the heterotic theory does not
depend on external metric.
Therefore if we include all the factors independent on
$E$ in $Z_R (m)/Z_R(\Lambda)$
into a normalization of $\psi^{(0)}$ then we can deduce one more
relation between
connections $c$ and $Z(\Lambda)/Z(m)$ and the correlator in Witten's theory.
Indeed the correlator in Witten's theory is given by an integral
of a (4,4) form $\Omega$ over ${\cal M} .$
Under a variation in metric of $M$ this form changes by an exact (4,4)
form $\bar{\partial}\omega + \partial \bar{\omega}$
($\partial$ and $\bar{\partial}$ are the Dolbeault \op s on ${\cal M}$
and $\omega$ and $\bar{\omega}$ are (3,4) and (4,3) forms on ${\cal M} ,$
respectively)
which in Witten's theory does not contribute to the correlator.
In the heterotic theory the physical correlator is given
essentially by an integral of
$\Omega \times (\det \hat{E}_R /\det \hat{E})^{1/2}$ over ${\cal M} .$
Because the correlator in the heterotic theory does not depend
on K\"ahlerian deformations of the metric we can assume that
the variation of the integrand should be exact too.
This gives us a certain equation which includes $c$ and some parameters
related to the geometry of zero modes in adjoint representation.

We discussed above a special class of physical correlators in the heterotic
model.
Of course we can choose a different physical \op \ for the sector of matter,
for example, the local \op \ constructed of the scalar field $\phi$
(however for the regulator path integral it is convenient to leave
the mass \op \ considered above).
In this case we get physical correlators which can not be obtained
from Witten's theory by the described procedure.
However since the anomaly in the BRST charge is cancelled
we can generalize the above discussion to this case.

It is to be emphasized that the physical correlator in the heterotic model
is given by an integral of a product of factors each of which is not in
general globally defined on ${\cal M} .$
However the whole correlator should be well defined because
the original SUSY theory seems to be well defined on K\"ahler manifold.

\section{Conclusions}

To conclude we have shown that N=1 SUSY gauge theory with an appropriate
representation of supermultiplet of matter can be twisted on K\"ahler
manifold $M.$
The twisted theory is shown to have a BRST charge $Q$ with a nontrivial
cohomology ring which determines the ring of physical \op s.
The lagrangian of such a heterotic model is $Q$-exact.
The physical correlators turn out to be independent on K\"ahlerian deformations
of metric (the dependence on the K\"ahler class of the
metric is factorized out into a power of a ratio
of two Ray-Singer torsions for the Dolbeault complex)
and on the gauge coupling constant.
Therefore they can be calculated in a semiclassical approximation
near anti-instanton configurations.

The correlators of the physical local operators
do not depend on anti-holomorphic coordinates
but they can depend on holomorphic coordinates on $M.$
However due to integration over the moduli space of instanton
these correlators can have singularities.
This is an interesting possibility because these correlators
could be interpreted as correlators in some two-dimensional theory.
Such an effective two-dimensional theory (if it exists)
is not pure topological in a sense that the physical correlators
can depend on holomorphic coordinates and therefore
can correspond to two-dimensional propagating degrees of freedom.

Thus the physical correlators are sections of a
holomorphic bundle on $M$ with coefficients which are topological invariants.
Actually the physical correlators depend on the complex structure
on $M.$
It is an open question if this dependence can be factorized out from
the dependence on smooth structure.

There is also an interesting possibility that such a twisting
induces a set of Ward identities for topological correlators
in an untwisted theory which could allow to calculate some of
the correlators in an untwisted theory in
terms of holomorphic sections of a vector bundle
on the K\"ahler manifold.
In particular the interesting problem is also whether
the heterotic topological theory allows to formulate
an analog of the $tt^*$ fusion equations \cite{vafa} for non-topological
amplitudes in SUSY theory in 4D.

As we have seen in the heterotic model some connection between
zero modes for adjoint and non-adjoint representations appears.
Therefore there is a
tempting possibility that such a heterotic topological theory can
be a tool for a study of zero modes of fields of matter in
non-adjoint represenation of the gauge group from the point of view of
geometry of the moduli space of instanton.

An open question is also if
the correlators in the heterotic topological model can be
interpreted in terms of $S$-matrix elements
in heterotic string theory.

\section{Acknowledgments}

I am very grateful to A.Lossev for collaborating in some part of this work,
for encouragements and interesting ideas.
I thank M.Bershadsky, M.Shifman, C.Vafa and A.Vainshtein for useful
discussions and comments.
I am very indebted to University of Minnesota, Fermi National Accelerator
Laboratory and  Niels Bohr Institute where a part of this work has been done
for warm hospitality.


\begin{thebibliography}{11}

\bibitem{vs} V.A.Novikov, M.A.Shifman, A.I.Vainshtein and V.I.Zakharov,
Nucl. Phys. {\bf B229} (1983) 407.
\bibitem{affleck} I.Affleck, M.Dine, and N.Seiberg, Nucl. Phys. {\bf B241}
(1984) 493; Nucl. Phys. {\bf B256} (1985) 557.
\bibitem{vs2} V.A.Novikov, M.A.Shifman, A.I.Vainshtein and V.I.Zakharov,
Nucl. Phys. {\bf B260} (1985) 157.
\bibitem{amati} D.Amati, K.Konishi, Y.Meurice, G.C.Rossi and G.Veneziano,
Phys. Rep. {\bf 162} (1988) 169, and references therein.
\bibitem{witten} E.Witten, Comm. Math. Phys. {\bf 117} (1988) 353.
\bibitem{witten2} E.Witten, Comm. Math. Phys. {\bf 118} (1988) 411.
\bibitem{eguchi} T.Eguchi and S.K.Yang, Mod. Phys. Lett. {\bf A4} (1990) 1693.
\bibitem{phase} E.Witten, Nucl. Phys. {\bf B340} (1990) 281.
\bibitem{ring} C.Vafa, Mod. Phys. Lett. {\bf A6} (1991) 337.
\bibitem{vafa} S.Cecotti and C.Vafa, Nucl. Phys. {\bf B367} (1991) 359.
\bibitem{pasq} S.Cecotti, L.Girardello and A.Pasquinucci, Int. J. Mod. Phys.
{\bf A6} (1991) 2427.
\bibitem{cec2} S.Cecotti, Int. J. Mod. Phys. {\bf A6} (1991) 1749;
Nucl. Phys. {\bf B 355} (1991) 755.
\bibitem{ym} J.P.Yamron, Phys. Lett. {\bf 213B} (1988) 325.
\bibitem{kr} A.Karlhede and M.Rocek, Phys. Lett. {\bf 212B} (1988) 51.
\bibitem{galperin} A.Galperin and O.Ogievetsky, Comm. Math. Phys.
\bibitem{don} S.Donaldson, Topology {\bf 29} (1990) 257.
{\bf 139} (1991) 377.
\bibitem{harvey} J.A.Harvey and A.Strominger, Comm. Math. Phys. {\bf 151}
(1993) 221.
\bibitem{west} M.F.Sohnius and P.C.West, Nucl. Phys. {\bf B198} (1981) 493;
Phys. Lett. {\bf 105B} (1981) 353.
\bibitem{ferrara} S.Ferrara, S.Sabharwal and M.Villasante, Phys.Lett.
{\bf 205B} (1988) 302.
\bibitem{gsw} M.B.Green, J.H.Schwarz and E.Witten,
Superstirng Theory, vol.2 (Cambridge Univ. Press, Cambridge,1987).
\bibitem{seiberg} N.Seiberg, Exact Results on the Space of Vacua of
Four Dimensional SUSY Gauge Theories. Preprint RU-94-18, hep-th/9402044.
\bibitem{ginsparg}P.Ginsparg, ``Applied conformal field theories'',
Les Houches Session XLIV, 1988, in Fields, Strings, and Critical
Phenomena, ed. by E.Brezin and J.Zinn-Justin, North Holland (1989).
\bibitem{freed} D.S.Freed and K.K.Uhlenbeck, Instantons and
four-manifolds, Springer,1984.
\bibitem{abc} A.Vainshtein, V.Zakharov, V.Novikov, and M.Shifman,
Uspekhi Fiz. Nauk, {\bf 136} (1982) 553.
\bibitem{ray} D.B.Ray and I.M.Singer, Advan.Math. {\bf 7} (1971) 145;
Ann. Math. {\bf 98} (1974) 154.
\bibitem{atiyah} M.F.Atiyah, F.R.S., N.J.Hitchin and I.M.Singer,
Proc. R. Soc. Lond. {\bf A 362} (1978) 425.
\bibitem{itoh} M.Itoh, J. Math. Soc. Jpn. {\bf 40} (1988) 9.

\end{thebibliography}
\end{document}